\documentclass[12pt]{article}
\usepackage{amsmath,amssymb}
\usepackage{bm}
\usepackage{color}
\usepackage{graphicx}
\usepackage{cite}
\usepackage{mathtools}
\usepackage{here}

\usepackage{multirow}
\usepackage {tabularx}
\newcolumntype{C}{>{\centering\arraybackslash}X}
\newcolumntype{L}{>{\raggedright\arraybackslash}X}
\newcolumntype{R}{>{\raggedleft\arraybackslash}X}

\renewcommand{\baselinestretch}{1.1}
\newcommand{\gfrac}[2]{\genfrac{}{}{0pt}{1}{#1}{#2}}
\newcommand{\tr}[1]{\hspace{.1em}\mathrm{tr\hspace{.1em}#1}}

\def\slash#1{\not\!\!#1}

\newcommand{\red}[1]{\textcolor{red}{#1}}
\begin{document}

\title{
\begin{flushright}
\ \\*[-80pt]
\begin{minipage}{0.2\linewidth}
\normalsize
EPHOU-23-022\\*[50pt]
\end{minipage}
\end{flushright}
{\Large \bf
CP phase in modular flavor models \\ and discrete Froggatt-Nielsen models
\\*[20pt]}}

\author{
Shota Kikuchi,
~Tatsuo Kobayashi, and
~Kaito Nasu
\\*[20pt]
\centerline{
\begin{minipage}{\linewidth}
\begin{center}
{\it \normalsize
Department of Physics, Hokkaido University, Sapporo 060-0810, Japan} \\*[5pt]
\end{center}
\end{minipage}}
\\*[50pt]}

\date{
\centerline{\small \bf Abstract}
\begin{minipage}{0.9\linewidth}
\medskip
\medskip
\small
We study the large mass hierarchy and CP violation in the modular symmetric quark flavor models without fine-tuning.
Mass matrices are written in terms of modular forms.
Modular forms near the modular fixed points are approximately given by $\varepsilon^p$, where $\varepsilon$ and $p$ denote the small deviation from the fixed points and their residual charges.
Thus mass matrices have the hierarchical structures depending on the residual charges, and have a possibility describing the large mass hierarchy without fine-tuning.
Similar structures of mass matrices are also obtained in Froggatt-Nielsen models.
Nevertheless, it seems to be difficult to induce a sufficient amount of CP violation by a single small complex parameter $\varepsilon$.
To realize the large mass hierarchy as well as sizable CP violation, multi-moduli are required.
We show the mass matrix structures with multi-moduli which are consistent with quark flavor observables including CP phase.
We also discuss the origins of the large mass hierarchy and CP violation in such mass matrix structures.
\end{minipage}
}

\begin{titlepage}
\maketitle
\thispagestyle{empty}
\end{titlepage}

\newpage


\section{Introduction}
\label{Intro}

The origin of the flavor structure is one of important issues to study in particle physics.
The experimental values of fermion masses show a large hierarchy.
For example, the ratio of the up quark mass to the top quark mass is very suppressed as 
$m_u/m_t ={\cal O}(10^{-5})$.
Other mass ratios are also suppressed.
Fermion mixing angles are also non-trivial.
The mixing angles are large in the lepton sector while they are small in the quark sector.
Furthermore, the origin of CP violation is unknown, although it may originate from 
underlying theory.
Underlying theory may be CP symmetric, and CP is broken spontaneous by some mechanism.
For example, higher dimensional theory such as string theory is CP symmetric 
\cite{Green:1987mn,Strominger:1985it,Dine:1992ya,Choi:1992xp,Lim:1990bp,Kobayashi:1994ks}, and 
CP may be violated by compactifiction.

In these years, modular flavor symmetries were studied intensively \cite{Feruglio:2017spp}.
(See for earlier works Refs.~~\cite{Kobayashi:2018vbk,Penedo:2018nmg,Criado:2018thu,Kobayashi:2018scp,Novichkov:2018ovf,Novichkov:2018nkm,deAnda:2018ecu,Okada:2018yrn,Kobayashi:2018wkl,Novichkov:2018yse} and for reviews Refs.~~\cite{Kobayashi:2023zzc,Ding:2023htn}.)
The modular symmetry is geometrical symmetry of compact space, and it would originate from string compactificaiton like 
torodical compactification and orbifold compactification \cite{Ferrara:1989qb,Lerche:1989cs,Lauer:1990tm,Kobayashi:2018rad,Kobayashi:2018bff,Ohki:2020bpo,Kikuchi:2020frp,Kikuchi:2020nxn,
Kikuchi:2021ogn,Almumin:2021fbk}.
Yukawa couplings and masses as well as other couplings are modular forms, which are holomorphic functions of the modulus $\tau$.
The modular symmetry is broken by fixing a value of the modulus, but 
residual $Z_N$ symmetries remain at fixed points.
Modular forms $Y(\tau)$, which can be Yukawa couplings, behave like $Y(\tau) \sim \varepsilon^p$ near a fixed point, where 
$p$ is a charge of residual $Z_N$ symmetry and $\varepsilon$ corresponds to a small deviation from the fixed point.
This behavior is useful to realize fermion mass hierarchies without fine-tuning \cite{Feruglio:2021dte,Novichkov:2021evw,Petcov:2022fjf,Kikuchi:2023cap,Abe:2023ilq,Kikuchi:2023jap,Abe:2023qmr,Petcov:2023vws,Abe:2023dvr,deMedeirosVarzielas:2023crv,Kikuchi:2023dow}.
In addition such a behavior is similar to one in Froggatt-Nielsen models \cite{Froggatt:1978nt}.

The modulus $\tau$ transforms as $\tau \to -\tau^*$ under CP in 
modular flavor models \cite{Baur:2019kwi,Novichkov:2019sqv,Baur:2019iai,Baur:2020jwc,Nilles:2020gvu}. 
Thus, the CP symmetry may be spontaneously broken by  a non-vanishing value of ${\rm Re }\tau$ 
except ${\rm Re }\tau = \pm 1/2$ and $|\tau|=1$, which are the boundary and the arc of the fundamental domain.
However, it seems to be difficult to realize sizable CP violation with realizing a large hierarchy of quark masses 
by starting from explicit CP-symmetric models with a single modulus \cite{Petcov:2022fjf,Kikuchi:2023cap,Abe:2023ilq,Kikuchi:2023jap,Abe:2023qmr,Petcov:2023vws,Abe:2023dvr,deMedeirosVarzielas:2023crv,Kikuchi:2023dow}.
On the other hand, we can realize sizable CP violation and a  large hierarchy of quark masses 
by multi-moduli \cite{Kikuchi:2023cap,Abe:2023ilq,Kikuchi:2023jap,Kikuchi:2023dow}.
Our purpose of this paper is to study what is an important point to realize a large mass hierarchy and sizable CP violation.

This paper is organized as follows.
In section \ref{sec:Froggatt-Nielsen models}, we give a brief review on $U(1)$ Froggatt-Nielsen models.
We examine the CP phase by the flavon field.
In section \ref{sec:Modular flavor models near fixed points}, we review modular flavor symmetric models and study the key point on CP violation by moduli.
In section \ref{sec:Quark mass matrices with two moduli parameters}, we study the quark mass matrices with two moduli parameters.
Section \ref{sec:conclusion} is our conclusion.
In Appendix \ref{app:patterns of charges}, we show patterns of $(T_1,T_2)$-charges and signs in coupling coefficients giving phenomenologically favorable mass matrix structures.


\section{$U(1)$ Froggatt-Nielsen models}
\label{sec:Froggatt-Nielsen models}

As said above, modular flavor models near fixed points have a behavior similar to Froggatt-Nielsen models.
Here, we briefly review about $U(1)$ Froggatt-Nielsen models.
Then, we clarify the key point on CP violation.


We discuss global supersymmetric $U(1)$ Froggatt-Nielsen models, 
because most of modular flavor symmetric models are 
studied within the framework of supersymmetric models.
We normalize $U(1)$ charge such that the flavon field $\phi$ has $U(1)$ charge $p=1$.
Its vacuum expectation value (VEV) $\langle \phi \rangle$ breaks the $U(1) $ flavor symmetry.
We denote $U(1)$ charges of three generations of quark doublets $Q_i$, right-handed up-sector quarks $u_i$, and 
right-handed down-sector quarks $d_i$ by $-p_{Qi}$, $-p_{ui}$, and $-p_{di}$, which are integers.
We assume that both the up-sector and down-sector Higgs fields, $H_u$ and $H_d$, have vanishing charge.
Then, effective Yukawa coupling terms in the superpotential are written by 
\begin{align}\label{eq:FN-Y}
a_{ij} \varepsilon^{p_{Qi}+p_{uj}}Q_iu_jH_u+ b_{ij} \varepsilon^{p_{Qi}+p_{dj}}Q_id_jH_d,
\end{align}
where $\varepsilon = \langle \phi \rangle/M$ and $M$ is a cut-off scale.
We expect that the coefficients
$a_{ij}$ and $b_{ij}$ are of ${\cal O}(1)$ and $\varepsilon$ is small, i.e. $|\varepsilon| \ll 1$.
One can realize quark mass ratios and mixing angles by choosing $U(1)$ charges and $\varepsilon$ properly \cite{Leurer:1992wg,Leurer:1993gy,Ibanez:1994ig,Binetruy:1994ru,Dudas:1995yu}.
For example, we assume 
\begin{align}
&p_{Q1}-p_{Q3}=3, \quad p_{Q2}-p_{Q3}=2,  \quad p_{u1}-p_{u3}=5,  \quad p_{u2}-p_{u3}=2,    \notag \\ 
&  p_{d1}-p_{d3}=1,  \quad p_{d2}-p_{d3}=0.
\end{align}
Then, up-sector and down-sector Yukawa matrices are obtained as
\begin{align}
Y_u \sim
\begin{pmatrix}
\varepsilon^8 & \varepsilon^5 &  \varepsilon^3 \\ 
\varepsilon^7 & \varepsilon^4 & \varepsilon^2\\
 \varepsilon^5 & \varepsilon^2 & 1\\
\end{pmatrix}, \qquad 
Y_d \sim
\begin{pmatrix}
\varepsilon^4 & \varepsilon^3 &  \varepsilon^3 \\ 
\varepsilon^3 & \varepsilon^2 & \varepsilon^2\\
 \varepsilon & 1 & 1\\
\end{pmatrix}, 
\end{align}
up to ${\cal O}(1)$ coefficients, $a_{ij}$ and $b_{ij}$.
They can lead to realistic values of quark masses and mixing angles 
when $\varepsilon \sim 0.2$.

We assume that underlying theory is CP symmetric, and 
all of coefficients, $a_{ij}$ and $b_{ij}$, are real.
The VEV $\langle \phi \rangle$ may violate CP.
We write $\varepsilon=|\varepsilon|e^{i \varphi}$.
However, we can rephase 
\begin{align}\label{eq:rephase}
Q_i \to e^{i p_{Qi} \varphi} Q_i, \qquad u_i \to e^{i p_{ui} \varphi} u_i, \qquad d_i \to e^{i p_{di} \varphi} d_i.
\end{align}
In such a field basis, the CP phases in mass matrices disappear.
Thus, there is no CP violation when all of coefficients, $a_{ij}$ and $b_{ij}$, are real.


\section{Modular flavor models near fixed points}
\label{sec:Modular flavor models near fixed points}

Here, we briefly review modular flavor symmetric models and 
study the key point on CP violation by moduli.

The modular group $SL(2,Z) \equiv \Gamma$ consists of $(2\times 2 )$ matrices $\gamma$,
\begin{align}
\gamma = 
\begin{pmatrix}
a & b \\  c & d
\end{pmatrix},
\end{align}
where $a,b,c,d$ are integers satisfying $ad-bc=1$.
The modular group is generated by $S$ and $T$,
\begin{align}
S=
\begin{pmatrix}
0 & 1 \\ -1 & 0
\end{pmatrix}, \quad
T=
\begin{pmatrix}
1 & 1 \\   0 & 1
\end{pmatrix}.
\end{align}
They satisfy the following algebraic relations:
\begin{align}
S^4=(ST)^3=1.
\end{align}

The modulus $\tau$ transforms  
\begin{align}
\tau \to \gamma \tau =\frac{a\tau +b}{c\tau + d},
\end{align}
under the modular symmetry.
In particular, the generators, $S$ and $T$, transform $\tau$ as 
\begin{align}
S: \tau \to -\frac{1}{\tau}, \qquad T: \tau \to \tau +1.
\end{align}
It is found that $S^2\tau = \tau$.
Thus, the modular group on $\tau$ corresponds to $PSL(2,Z)=\bar \Gamma$.
In addition, we define the principal congruence subgroup $\Gamma(M)$, 
\begin{align}
\Gamma(M) = \left\{  
\begin{pmatrix}
a & b \\ c & d 
\end{pmatrix}
\in \Gamma ~~\middle|~
\begin{pmatrix}
a & b \\ c & d 
\end{pmatrix}
=
\begin{pmatrix}
1 & 0 \\ 0 & 1 
\end{pmatrix}
\quad ({\rm mod}~M) 
\right\}   .
\end{align}

The modular forms $f_i(\tau)$ are holomorphic functions of $\tau$, which transform 
\begin{align}
f_i(\gamma \tau) = (c \tau +d)^k\rho_{ij}(\gamma)f_j(\tau),
\end{align}
under $\bar \Gamma$ ($\Gamma$), 
where $k$ is the modular weight and $\rho_{ij}(\gamma)$ is a unitary matrix.
Suppose that they transform 
\begin{align}
f_i(\gamma \tau) = (c \tau +d)^kf_i(\tau),
\end{align}
where $\gamma \in \bar \Gamma(M)$ ($\gamma \in  \Gamma(M)$).
Then, the unitary matrix $\rho_{ij}(\gamma)$ corresponds to a representation of 
quotient $\Gamma_M=\bar \Gamma/\bar \Gamma(M)$ ($\Gamma_M'= \Gamma/ \Gamma(M)$).
The quotients, $\Gamma_M$ with $M=2,3,4,5$ are isomorphic to $\Gamma_2 \approx S_3$, 
$\Gamma_3 \approx A_4$,  $\Gamma_4 \approx S_4$, $\Gamma_5 \approx A_5$ \cite{deAdelhartToorop:2011re}.
The quotients $\Gamma'_M =  \Gamma/ \Gamma(M)$ with $M=3,4,5$ are double covering 
groups of $\Gamma_M$.

Although the modular symmetry is broken by fixing the modulus value $\tau$, 
some residual $Z_N$ symmetries remain at fixed points of $\tau$.
The $Z_3$ symmetry generated by $ST$ remains at $\tau = e^{2\pi i/3}$.
The $Z_2$ and $Z_4$ symmetries generated by $S$ remain at $\tau = i$ in $\Gamma_M$ and $\Gamma'_M$, respectively.
In addition, $Z_M$ symmetry remains in the limit $\tau \to i \infty$ in $\Gamma_M$ and $\Gamma'_M$.

In modular flavor models, Yukawa couplings are written by modular forms $Y(\tau)$.
Thus, Yukawa couplings also transform non-trivially under the modular symmetry.
The modular form can be expanded by \cite{Novichkov:2021evw}
\begin{align}
Y(\tau) = \varepsilon^p + {\cal O}(\varepsilon^{p+N}),
\end{align}
near the fixed point, where the $Z_N$ symmetry remains.
Here, $\varepsilon$ denotes a small deviation from the fixed point, and 
$p$ is the $Z_N$ charge of the modular form.
That can lead to a flavor structure similar to the effective Yukawa couplings 
(\ref{eq:FN-Y})
in the Froggatt-Niesen models.

One assigns three-dimensional (reducible or irreducible) representations to three generations of quarks 
in $\Gamma_M$ and $\Gamma'_M$ modular flavor models.
Yukawa couplings are also their representations, because they are modular forms.
The superpotential must be invariant under $\Gamma_M$ and $\Gamma'_M$ including transformation behaviors of 
Yukawa couplings.

Here we focus on the residual $Z_N$ symmetry.
Suppose that three generations of quark doublets $Q_i$, right-handed up-sector quarks $u_i$, and 
right-handed down-sector quarks $d_i$ have the $Z_N$ charges as  $-p_{Qi}$, $-p_{ui}$, and $-p_{di}$.
Yukawa couplings must cancel their charges for the superpotential to be invariant.
When all $Z_N$ charges satisfy $0 \le p_{Qi}+p_{uj}< N$,  $0\le p_{Qi}+p_{dj}<N$, Yukawa couplings are written by 
\begin{align}
(Y_u)_{ij} =a_{ij}\varepsilon^{p_{Qi}+p_{uj}} + {\cal O}(\varepsilon^{N}), \qquad 
(Y_d)_{ij} =b_{ij}\varepsilon^{p_{Qi}+p_{dj}}+{\cal O}(\varepsilon^{N}),
\end{align}
where $a_{ij}$ and $b_{ij}$ are ${\cal O}(1)$ coefficients, and they may be related each other 
when quark fields $Q_i$, $u_i$, $d_i$ are transformed as irreducible multiplets of finite modular groups.  
The dominant terms are the same form as Yukawa couplings in the superpotential (\ref{eq:FN-Y}).
Hence, the phase of $\varepsilon$ can be rephased by Eq.(\ref{eq:rephase}), and there remains no physcial CP phase.
The subdominant terms are small when $\varepsilon = {\cal O}(0.1)$ and $N$ is large, e.g. $N>5$.
Then, the sizable CP violation can not be realized.

The situation is different when some  of ${\rm mod}(p_{Qi}+p_{uj},N)$ and  ${\rm mod}(p_{Qi}+p_{dj},N)$ are different from others, 
where mod$(p,N)=r$ for integers $p,q,r$ satisfying $p=qN+r$, $N>r\geq0$.
For example, suppose that $p_{Q1}+p_{u1}= N$ and the others satisfy 
$0 \le p_{Qi}+p_{uj}< N$,  $0\le p_{Qi}+p_{dj}<N$.
Then, Yukawa couplings are written by 
\begin{align}
Y_u\sim 
\begin{pmatrix}
a_{11} & a_{12}\varepsilon^{p_{Q1}+p_{u2}} & a_{13}\varepsilon^{p_{Q1}+p_{u3}} \\
a_{21}\varepsilon^{p_{Q2}+p_{u1}} & a_{22}\varepsilon^{p_{Q2}+p_{u2}} & a_{23}\varepsilon^{p_{Q2}+p_{u3}} \\
a_{31}\varepsilon^{p_{Q3}+p_{u1}} & a_{32}\varepsilon^{p_{Q3}+p_{u2}} & a_{33}\varepsilon^{p_{Q3}+p_{u3}}
\end{pmatrix},
\notag \\
Y_d\sim 
\begin{pmatrix}
b_{11} \varepsilon^{p_{Q1}+p_{d1}}& b_{12}\varepsilon^{p_{Q1}+p_{d2}} & b_{13}\varepsilon^{p_{Q1}+p_{d3}} \\
b_{21}\varepsilon^{p_{Q2}+p_{d1}} & b_{22}\varepsilon^{p_{Q2}+p_{d2}} & b_{23}\varepsilon^{p_{Q2}+p_{d3}} \\
b_{31}\varepsilon^{p_{Q3}+p_{d1}} & b_{32}\varepsilon^{p_{Q3}+p_{d2}} & b_{33}\varepsilon^{p_{Q3}+p_{d3}}
\end{pmatrix}.
\end{align}
We can not rephase such that the CP phase disappears in this case, 
but a CP phase remain.
However, experimental values of quark mass hierarchies as well as mixing angles can not be realized properly.
We have a similar problem in the models, where 
some of ${\rm mod}(p_{Qi}+p_{uj},N)$ and  ${\rm mod}(p_{Qi}+p_{dj},N)$ are different from others.
Thus, it is difficult to realize a sizable CP phase and the proper hierarchy of quark  masses and mixing angles 
at the same time by a single modulus VEV in the modular flavor models 
as well as a single flavon VEV in the $Z_N$ Froggatt-Nielsen models.

In the next section, we extend our analysis to the modular flavor models with multi-moduli.
For simplicity, we study the models with two moduli, $\tau_1$ and $\tau_2$, which have the 
modular symmetry $\Gamma_{M_1}^{(')} \times \Gamma_{M_2}^{(')}$.
We focus on their residual symmetries at fixed points, $Z_{N_1} \times Z_{N_2}$.
Our models are equivalent to $Z_{N_1} \times Z_{N_2}$ Froggatt-Nielsen models.


\section{Quark mass matrices with two moduli parameters}
\label{sec:Quark mass matrices with two moduli parameters}

In this section, we study the structures of the quark mass matrices with two moduli parameters $(\tau_1,\tau_2)$.
We assume the moduli near the modular fixed point for both $T_1$ and $T_2$-transformations, $(\tau_1,\tau_2)\sim (i\infty, i\infty)$, where residual $Z_{N_1}$ and $Z_{N_2}$-symmetries remain.
Then mass matrices become hierarchical depending on their residual $(T_1,T_2)$-charges.
In the following, we classify $(T_1,T_2)$-charges of $Q_i$, $u_i$ and $d_i$ leading to phenomenologically favorable structures of quark mass matrices.
Note that our analysis in this section can be applied for discrete Froggatt-Nielsen models with two flavon VEV parameters as mentioned in previous section.

For simplicity, we assume that residual $Z_{N_1}$ and $Z_{N_2}$-symmetries correspond to $T_1$ and $T_2$-transformations.
When $N_1$ is small like $N_1=2,3,4$, the residual $Z_{N1}$ symmetry can originate from $S_1$ and $S_1T_1$ symmetries 
at $\tau_1= i$ and $e^{2 \pi i/3}$.
Even when $N_2$ is large like $N_2=6,8,9$, the residual $Z_{N_2}$ symmetry can originate from a proper combination of 
two residual symmetries, i.e., $Z_{N_2}=Z_{N_3} \times Z_{N_4}$, where $N_3, N_4=2,3,4$.


\subsection{Phenomenologically favorable structures}

Yukawa couplings with $(T_1, T_2)$-charges $(m,n)$ can be approximated as
\begin{align}
Y(\tau_1,\tau_2) \sim \varepsilon_1^{m} \cdot \varepsilon_2^{n},
\quad m \in Z_{N_1}, \quad n \in Z_{N_2},
\end{align}
where $\varepsilon_1\equiv e^{2\pi i\tau_1/N_1}$ and $\varepsilon_2\equiv e^{2\pi i\tau_2/N_2}$.
Here we denote $(T_1, T_2)$-charges of $Q_i$, $u_i$ and $d_i$ as $(-p_{Qi1}, -p_{Qi2})$, $(-p_{ui1}, -p_{ui2})$ and $(-p_{di1}, -p_{di2})$, respectively.
Also we assume that both the up-sector and down-sector Higgs fields have vanishing charge.
Then the mass matrix elements of the up-sector and down-sector quarks in the first order approximation of $\varepsilon_1$ and $\varepsilon_2$ are written as
\begin{align}
&M_u^{ij} = 
a_{ij}\varepsilon_1^{\textrm{mod}(p_{Qi1}+p_{uj1},N_1)} \cdot \varepsilon_2^{\textrm{mod}(p_{Qi2}+p_{uj2},N_2)}
\langle H_u \rangle, \\
&M_d^{ij} = 
b_{ij}\varepsilon_1^{\textrm{mod}(p_{Qi1}+p_{dj1},N_1)} \cdot \varepsilon_2^{\textrm{mod}(p_{Qi2}+p_{dj2},N_2)}
\langle H_d \rangle,
\end{align}
where $a_{ij}$ and $b_{ij}$ are ${\cal O}(1)$ coefficients and they may be related each other when $Q_i$, $u_i$, and  $d_i$ are transformed as multiplets of finite modular groups.
Note that the residual charges of the mass matrix elements are determined to cancel ones of $Q_i$, $u_i$ and $d_i$.
We assume $(T_1,T_2)$-charges such that $M_u^{33}$ and $M_d^{33}$ have vanishing charges,
\begin{align}
\begin{aligned}
&(T_1,T_2)~\textrm{of}~M_u^{33} = (\textrm{mod}(p_{Q31}+p_{u31},N_1), \textrm{mod}(p_{Q32}+p_{u32},N_2)) = (0,0), \\
&(T_1,T_2)~\textrm{of}~M_d^{33} = (\textrm{mod}(p_{Q31}+p_{d31},N_1), \textrm{mod}(p_{Q32}+p_{d32},N_2)) = (0,0).
\end{aligned} \label{eq:vanishing33}
\end{align}
Hence, we assume $M_u^{33}=a^{33}\langle H_u\rangle$ and $M_d^{33}=b^{33}\langle H_d\rangle$.
We adopt $p_{Q31}=p_{Q32}=p_{u31}=p_{u32}=p_{d31}=p_{d32}=0$ as $(T_1,T_2)$-charges satisfying Eq.~(\ref{eq:vanishing33}).
Contrary to the mass matrices with single modulus parameter, it can be expected that both large quark mass hierarchies and the CP violation are led from above mass matrices.
For instance, we can expect that large quark mass hierarchies originate from $\varepsilon_1$ while the CP violation originates from $\varepsilon_2$.
There may exist further possibilities leading to realistic values of them simultaneously.
In the following, we explore the structures of quark mass matrices with
\begin{itemize}
\item $(N_1,N_2)$ = (2,5), (2,6), (2,7), (2,8), (2,9), (3,4), (3,5), (3,6), (4,4), (4,5), (4,6);
\item $|\varepsilon_1|=|\varepsilon_2|^k$, $k=1/2, 1, 2$.
\end{itemize}
We call each choice of $(N_1,N_2,k)$ as ``type''.
There are different mass matrix structures depending on $(T_1,T_2)$-charges of $Q_i$, $u_i$ and $d_i$ in each type.
In each pattern of $(T_1,T_2)$-charges, we calculate the quark mass ratios, the absolute values of the Cabibbo-Kobayashi-Maskawa (CKM) matrix elements and the Jarlskog invariant and classify phenomenologically favorable models.
First we choose the benchmark point $x_{bp}\equiv |\varepsilon_1|=|\varepsilon_2|^k$ satisfying
\begin{align}
|M_u^{11}/M_u^{33}| = |\varepsilon_1|^{\textrm{mod}(p_{Q11}+p_{u11},N_1)} \cdot |\varepsilon_2|^{\textrm{mod}(p_{Q12}+p_{u12},N_2)} = (m_u/m_t),
\end{align}
in order to realize the ratio of the up quark mass to the top quark mass.
In addition, we introduce $|\varepsilon|$ such that $|\varepsilon|^6=(m_u/m_t)$.
Using $|\varepsilon|$, we can find $n_c$, $n_d$ and $n_s$ such that
\begin{align}
\begin{aligned}
&|M_u^{22}/M_u^{33}| = |\varepsilon_1|^{\textrm{mod}(p_{Q21}+p_{u21},N_1)} \cdot |\varepsilon_2|^{\textrm{mod}(p_{Q22}+p_{u22},N_2)} = |\varepsilon|^{n_c}, \\
&|M_d^{11}/M_d^{33}| = |\varepsilon_1|^{\textrm{mod}(p_{Q11}+p_{d11},N_1)} \cdot |\varepsilon_2|^{\textrm{mod}(p_{Q12}+p_{d12},N_2)} = |\varepsilon|^{n_d}, \\
&|M_d^{22}/M_d^{33}| = |\varepsilon_1|^{\textrm{mod}(p_{Q21}+p_{d21},N_1)} \cdot |\varepsilon_2|^{\textrm{mod}(p_{Q22}+p_{d22},N_2)} = |\varepsilon|^{n_s}.
\label{eq:powers_nc_nd_ns}
\end{aligned}
\end{align}
In terms of $|\varepsilon|$, the diagonal elements of $M_u$ and $M_d$ are rewritten as
\begin{align}
&M_u = 
\begin{pmatrix}
|\varepsilon|^6 & \ast & \ast \\
\ast & |\varepsilon|^{n_c} & \ast \\
\ast & \ast & 1 \\
\end{pmatrix} \langle H_u \rangle, \quad
M_u = 
\begin{pmatrix}
|\varepsilon|^{n_d} & \ast & \ast \\
\ast & |\varepsilon|^{n_s} & \ast \\
\ast & \ast & 1 \\
\end{pmatrix} \langle H_d \rangle.
\end{align}
To obtain realistic quark mass hierarchies, we introduce the constraints on powers $n_c$, $n_d$ and $n_s$.
We show the quark mass ratios at GUT scale with $\tan\beta=5$ in Table \ref{tab:quark_masses_GUT}.
\begin{table}[H]
\centering
\begin{tabular}{cccc} \hline
$\frac{m_u}{m_t}\times 10^6$ & $\frac{m_c}{m_t}\times 10^3$ & $\frac{m_d}{m_b}\times 10^4$ & $\frac{m_s}{m_b}\times 10^2$ \\ \hline
5.39 & 2.80 & 9.21 & 1.82 \\ \hline
\end{tabular}
\caption{Quark mass ratios at the GUT scale $2\times 10^{16}$ GeV with $\tan \beta=5$ \cite{Antusch:2013jca,Bjorkeroth:2015ora}.}
\label{tab:quark_masses_GUT}
\end{table}
These values indicate
\begin{align}
n_c \sim 3, \quad n_d \gtrsim 3, \quad n_s \sim 2.
\end{align}
To realize this situation, we impose following conditions,
\begin{align}
4 \geq n_d \geq n_c > 2, \quad n_c-1 \geq n_s > 1. \label{eq:conditions_of_powers}
\end{align}
Moreover, we need to obtain non-vanishing CP phase.
To understand the origin of CP violation, let us consider the phase transformations $M_u\to U_QM_u U_u$ and $M_d\to U_Q M_d U_d$, where
\begin{align}
&U_Q^{ij} = e^{-2\pi ip_{Qj1}\textrm{Re}\tau_1/N_1} e^{-2\pi ip_{Qj2}\textrm{Re}\tau_2/N_2} \delta_{i,j}, \\
&U_u^{ij} = e^{-2\pi ip_{uj1}\textrm{Re}\tau_1/N_1} e^{-2\pi ip_{uj2}\textrm{Re}\tau_2/N_2} \delta_{i,j}, \\
&U_d^{ij} = e^{-2\pi ip_{dj1}\textrm{Re}\tau_1/N_1} e^{-2\pi ip_{dj2}\textrm{Re}\tau_2/N_2} \delta_{i,j}.
\end{align}
After the phase transformations, we obtain
\begin{align}
&[U_Q M_u U_u]^{ij} = e^{2\pi i\phi^{ij}_{u1}\textrm{Re}\tau_1/N_1} e^{2\pi i\phi^{ij}_{u2}\textrm{Re}\tau_2/N_2} |\varepsilon_1|^{\textrm{mod}(p_{Qi1}+p_{uj1},N_1)} \cdot |\varepsilon_2|^{\textrm{mod}(p_{Qi2}+p_{uj2},N_2)}, \\
&[U_Q M_d U_d]^{ij} = e^{2\pi i\phi^{ij}_{d1}\textrm{Re}\tau_1/N_1} e^{2\pi i\phi^{ij}_{d2}\textrm{Re}\tau_2/N_2} |\varepsilon_1|^{\textrm{mod}(p_{Qi1}+p_{dj1},N_1)} \cdot |\varepsilon_2|^{\textrm{mod}(p_{Qi2}+p_{dj2},N_2)},
\end{align}
where
\begin{align}
&\phi_{u1}^{ij} = \textrm{mod}(p_{Qi1}+p_{uj1},N_1) - (\textrm{mod}(p_{Qi1},N_1) + \textrm{mod}(p_{uj1},N_1)), \\
&\phi_{u2}^{ij} = \textrm{mod}(p_{Qi2}+p_{uj2},N_2) - (\textrm{mod}(p_{Qi2},N_2) + \textrm{mod}(p_{uj2},N_2)), \\
&\phi_{d1}^{ij} = \textrm{mod}(p_{Qi1}+p_{dj1},N_1) - (\textrm{mod}(p_{Qi1},N_1) + \textrm{mod}(p_{dj1},N_1)), \\
&\phi_{d2}^{ij} = \textrm{mod}(p_{Qi2}+p_{dj2},N_2) - (\textrm{mod}(p_{Qi2},N_2) + \textrm{mod}(p_{dj2},N_2)).
\end{align}
Then the up-sector quark mass matrix element $[U_Q M_u U_u]^{ij}$ includes the phase factor $e^{-2\pi i\textrm{Re}\tau_1}$ when
\begin{align}
&\textrm{mod}(p_{Qi1},N_1) + \textrm{mod}(p_{uj1},N_1) \geq N_1, \label{eq:CP1}
\end{align}
is satisfied.
Similarly, $[U_Q M_u U_u]^{ij}$ includes the phase factor $e^{-2\pi i\textrm{Re}\tau_2}$ when
\begin{align}
&\textrm{mod}(p_{Qi2},N_2) + \textrm{mod}(p_{uj2},N_2) \geq N_2, \label{eq:CP2}
\end{align}
is satisfied.
We can find the same conditions for down-sector quarks,
\begin{align}
&\textrm{mod}(p_{Qi1},N_1) + \textrm{mod}(p_{dj1},N_1) \geq N_1, \label{eq:CP3} \\
&\textrm{mod}(p_{Qi2},N_2) + \textrm{mod}(p_{dj2},N_2) \geq N_2. \label{eq:CP4}
\end{align}
Thus we require either of these conditions as the necessary conditions for CP violation.

Our purpose is to find the mass matrix structures leading to the orders of the quark flavor observables without fine-tuning.
We expect that the values of coupling coefficients $a_{ij}$ and $b_{ij}$ are not origin of the large quark mass hierarchies.
It is natural to assume ${\cal O}(1)$ values of $a_{ij}$ and $b_{ij}$. 
Here, we restrict the values of coupling constants $a_{ij}$ and $b_{ij}$ to $\pm 1$ in order to avoid fine-tuinng by them.
We can choose the basis of fields such that signs in $a_{ij}$ and $b_{ij}$ are restricted to
\begin{align}
\begin{pmatrix}
a_{11} & a_{12} & a_{13} \\
a_{21} & a_{22} & a_{23} \\
a_{31} & a_{32} & a_{33} \\
\end{pmatrix}
=
\begin{pmatrix}
1 & 1 & 1 \\
1 & \pm 1 & \pm 1 \\
1 & \pm 1 & \pm 1 \\
\end{pmatrix}, \quad
\begin{pmatrix}
b_{11} & b_{12} & b_{13} \\
b_{21} & b_{22} & b_{23} \\
b_{31} & b_{32} & b_{33} \\
\end{pmatrix}
=
\begin{pmatrix}
1 & 1 & 1 \\
\pm 1 & \pm 1 & \pm 1 \\
\pm 1 & \pm 1 & \pm 1 \\
\end{pmatrix}.
\end{align}
Then we give numerical calculations of the quark mass ratios and the absolute values of the CKM matrix elements at CP symmetric point $(\varepsilon_1,\varepsilon_2)=(e^{2\pi i\textrm{Re}\tau_1/N_1}x_{bp}, e^{2\pi i\textrm{Re}\tau_2/N_2}x_{bp}^{1/k})=(x_{bp},x_{bp}^{1/k})$ to evaluate the orders of them \footnote{CP symmetry is not violated in $\textrm{Re}\tau_1=0$ and $\textrm{Re}\tau_2=0$.}.
For our purpose, we require the following conditions for the obtained values,
\begin{align}
\begin{aligned}
&1/3 < \frac{(m_u/m_t)_\textrm{obtained}}{(m_u/m_t)_\textrm{GUT}} < 3, \quad
1/3 < \frac{(m_c/m_t)_\textrm{obtained}}{(m_c/m_t)_\textrm{GUT}} < 3, \\
&1/3 < \frac{(m_d/m_b)_\textrm{obtained}}{(m_d/m_b)_\textrm{GUT}} < 3, \quad
1/3 < \frac{(m_s/m_b)_\textrm{obtained}}{(m_s/m_b)_\textrm{GUT}} < 3, \\
&2/3 < \frac{|V_\textrm{CKM}^x|_\textrm{obtained}}{|V_\textrm{CKM}^x|_\textrm{GUT}} < 3/2, \quad (x\in \{us,cb,ub\}). \label{eq:conditions_mass_CKM}
\end{aligned}
\end{align}
Note that we do not consider CP violation in above conditions.
This is because suitable benchmark points of the phase factors of $\varepsilon_1$ and $\varepsilon_2$, that is, $\textrm{Re}\tau_1$ and $\textrm{Re}\tau_2$, are non-trivial.
It is difficult to evaluate the order of CP phase (Jarlskog invariant) by numerical calculations.
However, there may exist the mass matrix structures not leading to sufficient CP violation even if the necessary conditions for the CP violation are satisfied.

Instead of numerical calculation, we estimate the order of Jarlskog invariant through the expansion of quark mass matrices by $x=|\varepsilon_1|\ll 1$, $p_1=e^{-2\pi i\textrm{Re}\tau_1}$ and $p_2=e^{-2\pi i\textrm{Re}\tau_2}$.
We can find the leading terms of quark flavor observables including Jarlskog invariant in the expansion.
Then we require non-vanishing Jarlskog invariant at the leading term to generate sufficient CP violation.

Here we summarize the conditions to find the mass matrix structures leading to the orders of the quark flavor observables.
\begin{enumerate}
\item The conditions for hierarchical masses in Eq.~(\ref{eq:conditions_of_powers}),
\begin{align}
4 \geq n_d \geq n_c > 2, \quad n_c-1 \geq n_s > 1. \notag
\end{align}
The powers $n_c$, $n_d$ and $n_s$ defined by Eq.~(\ref{eq:powers_nc_nd_ns}) must satisfy these conditions.
\item The necessary conditions for CP violation in Eqs.~(\ref{eq:CP1}), (\ref{eq:CP2}), (\ref{eq:CP3}) and (\ref{eq:CP4}),
\begin{align}
&\textrm{mod}(p_{Qi1},N_1) + \textrm{mod}(p_{uj1},N_1) \geq N_1, \notag \\
&\textrm{mod}(p_{Qi2},N_2) + \textrm{mod}(p_{uj2},N_2) \geq N_2, \notag \\
&\textrm{mod}(p_{Qi1},N_1) + \textrm{mod}(p_{dj1},N_1) \geq N_1, \notag \\
&\textrm{mod}(p_{Qi2},N_2) + \textrm{mod}(p_{dj2},N_2) \geq N_2. \notag
\end{align}
To obtain quark mass matrices with non-vanishing phase factors, either of these conditions are required.
\item The conditions to realize the orders of the quark mass ratios and the absolute values of the CKM matrix elements in Eq.~(\ref{eq:conditions_mass_CKM}).
Obtained values in numerical calculations at $(\varepsilon_1,\varepsilon_2)=(x_{bp}, x_{bp}^{1/k})$ must satisfy these conditions.
\item The condition to realize the order of Jarlskog invariant. \\
To induce sufficient CP violation, Jarlskog invariant must not vanish at the leading term in the expansion by $x=|\varepsilon_1|\ll 1$, $p_1=e^{-2\pi i\textrm{Re}\tau_1}$ and $p_2=e^{-2\pi i\textrm{Re}\tau_2}$.
Note that this is the sufficient condition of the necessary conditions for CP violation in Eqs.~(\ref{eq:CP1}), (\ref{eq:CP2}), (\ref{eq:CP3}) and (\ref{eq:CP4}).
On the other hand, we need the necessary conditions for CP violation for the efficiency of analysis.
\end{enumerate}
We regard mass matrix structures satisfying all conditions as phenomenologically favorable structures.

Consequently, we find the phenomenologically favorable structures in seven types $(N_1,N_2,k)=(2,7,\frac{1}{2})$, $(2,8,\frac{1}{2})$, $(2,8,1)$, $(2,9,\frac{1}{2})$, $(2,9,1)$, $(3,6,\frac{1}{2})$ and $(3,6,2)$.
We show the number of patterns of $(T_1,T_2)$-charges of $Q_i$, $u_i$ and $d_i$ and signs in coupling coefficients giving favorable structures in Table \ref{tab:number of models leading to J_CP}.

\begin{table}[H]
\centering
\begin{tabular}{c|ccccccccccc} \hline
$(N_1,N_2)$ & (2,5) & (2,6) & (2,7) & (2,8) & (2,9) & (3,4) & (3,5) & (3,6) & (4,4) & (4,5) & (4,6) \\ \hline
$k=\frac{1}{2}$ ($|\varepsilon_1|>|\varepsilon_2|$) & 0 & 0 & 6 & 154 & 268 & 0 & 0 & 18 & 0 & 0 & 0 \\
$k=1$ ($|\varepsilon_1|=|\varepsilon_2|$) & 0 & 0 & 0 & 26 & 70 & 0 & 0 & 0 & 0 & 0 & 0 \\
$k=2$ ($|\varepsilon_1|<|\varepsilon_2|$) & 0 & 0 & 0 & 0 & 0 & 0 & 0 & 16 & - & 0 & 0 \\ \hline
\end{tabular}
\caption{Number of patterns of $(T_1,T_2)$-charges of $Q_i$, $u_i$ and $d_i$ and signs in coupling coefficients giving favorable structures.
Here $k$ is defined by the relation $|\varepsilon_1|=|\varepsilon_2|^k$.}
\label{tab:number of models leading to J_CP}
\end{table}

In Tables \ref{tab:(2,7,1/2)}, \ref{tab:(2,8,1/2)}, \ref{tab:(2,8,1)}, \ref{tab:(2,9,1/2)}, \ref{tab:(2,9,1)}, \ref{tab:(3,6,1/2)} and \ref{tab:(3,6,2)}, we show the leading terms of observables obtained from favorable structures.
In columns of ``Observables'', we show the leading terms of observables in ordering shown in Table \ref{tab:observables}.
\begin{table}[H]
\centering
\begin{tabular}{c} \hline
Observables \\ \hline
$(\frac{m_u}{m_t}, \frac{m_c}{m_t})$ \\
$(\frac{m_d}{m_b}, \frac{m_s}{m_b})$ \\
$|V_{\textrm{CKM}}^{us}|$ \\
$|V_{\textrm{CKM}}^{cb}|$ \\
$|V_{\textrm{CKM}}^{ub}|$ \\
$J_{\textrm{CP}}=|\textrm{Im}(V_{\textrm{CKM}}^{us}V_{\textrm{CKM}}^{cb}(V_{\textrm{CKM}}^{ub}V_{\textrm{CKM}}^{cs})^*)|$ \\ \hline
\end{tabular}
\caption{Ordering of observables in Tables \ref{tab:(2,7,1/2)}, \ref{tab:(2,8,1/2)}, \ref{tab:(2,8,1)}, \ref{tab:(2,9,1/2)}, \ref{tab:(2,9,1)}, \ref{tab:(3,6,1/2)} and \ref{tab:(3,6,2)},.}
\label{tab:observables}
\end{table}

In Appendix \ref{app:patterns of charges}, we also show patterns of $(T_1,T_2)$-charges of $Q_i$, $u_i$ and $d_i$ and signs in coupling coefficients giving favorable structures.

\begin{table}[H]
\centering
\begin{tabular}{c|c|c||c|c|c} \hline
Label & \multirow{2}{*}{Observables} & \multirow{2}{*}{Values at $x_{bp}$} & Label & \multirow{2}{*}{Observables} & \multirow{2}{*}{Values at $x_{bp}$} \\
($x_{bp}$) & & & ($x_{bp}$) & & \\ \hline
{\scriptsize $\begin{array}{c}(2,7,\frac{1}{2})\bm{1}\\ (0.393)\\ \end{array}$} & $\begin{smallmatrix}
(2x^{13}, 2x^{8}) \\
(x^{8}, 2x^{5}) \\
0.5x \\ 2x^{4} \\ |p_1^{-1}x^{7} +2x^{7}| \\
2x^{12}|\textrm{Im}(p_1)| \\
\end{smallmatrix}$ & $\begin{smallmatrix}
( 1.07\times 10^{-5},  1.14\times 10^{-3}) \\
( 5.69\times 10^{-4},  1.87\times 10^{-2}) \\
0.196 \\ 0.0477 \\ |0.00145p_1^{-1} +0.00290| \\
0.0000271|\textrm{Im}(p_1)| \\
\end{smallmatrix}$ &
{\scriptsize $\begin{array}{c}(2,7,\frac{1}{2})\bm{2}\\ (0.393)\\ \end{array}$} & $\begin{smallmatrix}
(2x^{13}, 2x^{8}) \\
(2x^{8}, 2x^{5}) \\
0.5x \\ 2x^{4} \\ |p_1^{-1}x^{7} +2x^{7}| \\
2x^{12}|\textrm{Im}(p_1)| \\
\end{smallmatrix}$ & $\begin{smallmatrix}
( 1.07\times 10^{-5},  1.14\times 10^{-3}) \\
( 1.14\times 10^{-3},  1.87\times 10^{-2}) \\
0.196 \\ 0.0477 \\ |0.00145p_1^{-1} +0.00290| \\
0.0000271|\textrm{Im}(p_1)| \\
\end{smallmatrix}$ \\ \hline
\end{tabular}
\caption{The leading terms of observables obtained from favorable structures in $(N_1,N_2,k)=(2,7,\frac{1}{2})$.
The second and fifth columns show the structures of observables in the leading order approximation of $x\equiv|\varepsilon_1|=|\varepsilon_2|^k$.
The third and sixth columns show values of those at $x=x_{bp}$.
There are two different leading terms of observables labeled as $(2,7,\frac{1}{2})\bm{n}$ $(\bm{n}=\bm{1},\bm{2})$.}
\label{tab:(2,7,1/2)}
\end{table}

\begin{table}[H]
\caption{The leading terms of observables obtained from favorable structures in $(N_1,N_2,k)=(2,8,\frac{1}{2})$.
The second and fifth columns show the structures of observables in the leading order approximation of $x\equiv|\varepsilon_1|=|\varepsilon_2|^k$.
The third and sixth columns show values of those at $x=x_{bp}$.
There are 42 different leading terms of observables labeled as $(2,8,\frac{1}{2})\bm{n}$ $(\bm{n}=\bm{1},\bm{2},...,\bm{42})$.}
\label{tab:(2,8,1/2)}
\centering
\begin{tabular}{c|c|c||c|c|c} \hline
Label & \multirow{2}{*}{Observables} & \multirow{2}{*}{Values at $x_{bp}$} & Label & \multirow{2}{*}{Observables} & \multirow{2}{*}{Values at $x_{bp}$} \\
($x_{bp}$) & & & ($x_{bp}$) & & \\ \hline
{\scriptsize $\begin{array}{c}(2,8,\frac{1}{2})\bm{1}\\ (0.420)\\ \end{array}$} & $\begin{smallmatrix}
(4x^{16}, x^{8}) \\
(2x^{8}, x^{4}) \\
4x^{3} \\ |p_1^{-1}x^{5} +2x^{5}| \\  2x^{6} \\
16x^{14}|\textrm{Im}(p_1)| \\
\end{smallmatrix}$ & $\begin{smallmatrix}
( 3.75\times 10^{-6},  9.68\times 10^{-4}) \\
( 1.94\times 10^{-3},  3.11\times 10^{-2}) \\
0.296 \\ |0.0131p_1^{-1} +0.0261| \\ 0.0110 \\
0.0000850|\textrm{Im}(p_1)| \\
\end{smallmatrix}$ &
{\scriptsize $\begin{array}{c}(2,8,\frac{1}{2})\bm{2}\\ (0.420)\\ \end{array}$} & $\begin{smallmatrix}
(4x^{16}, x^{8}) \\
(4x^{10}, x^{4}) \\
4x^{3} \\ |p_1^{-1}x^{5} +2x^{5}| \\  2x^{6} \\
16x^{14}|\textrm{Im}(p_1)| \\
\end{smallmatrix}$ & $\begin{smallmatrix}
( 3.75\times 10^{-6},  9.68\times 10^{-4}) \\
( 6.83\times 10^{-4},  3.11\times 10^{-2}) \\
0.296 \\ |0.0131p_1^{-1} +0.0261| \\ 0.0110 \\
0.0000850|\textrm{Im}(p_1)| \\
\end{smallmatrix}$ \\
\hline
\end{tabular}
\end{table}
\begin{table}[H]
\centering
\begin{tabular}{c|c|c||c|c|c} \hline
Label & \multirow{2}{*}{Observables} & \multirow{2}{*}{Values at $x_{bp}$} & Label & \multirow{2}{*}{Observables} & \multirow{2}{*}{Values at $x_{bp}$} \\
($x_{bp}$) & & & ($x_{bp}$) & & \\ \hline
{\scriptsize $\begin{array}{c}(2,8,\frac{1}{2})\bm{5}\\ (0.420)\\ \end{array}$} & $\begin{smallmatrix}
(2x^{14}, x^{8}) \\
(4x^{10}, x^{4}) \\
 4x^{3} \\ |p_1^{-1}x^{5} +2x^{5}| \\ 2x^{6} \\
16x^{14}|\textrm{Im}(p_1)| \\
\end{smallmatrix}$ & $\begin{smallmatrix}
( 1.06\times 10^{-5},  9.68\times 10^{-4}) \\
( 6.83\times 10^{-4},  3.11\times 10^{-2}) \\
0.296 \\ |0.0131p_1^{-1} +0.0261| \\ 0.0110 \\
0.0000850|\textrm{Im}(p_1)| \\
\end{smallmatrix}$ &
{\scriptsize $\begin{array}{c}(2,8,\frac{1}{2})\bm{6}\\ (0.420)\\ \end{array}$} & $\begin{smallmatrix}
(2x^{14}, x^{8}) \\
(2x^{8}, x^{4}) \\
 4x^{3} \\ |p_1^{-1}x^{5} +2x^{5}| \\ 2x^{6} \\
16x^{14}|\textrm{Im}(p_1)| \\
\end{smallmatrix}$ & $\begin{smallmatrix}
( 1.06\times 10^{-5},  9.68\times 10^{-4}) \\
( 1.94\times 10^{-3},  3.11\times 10^{-2}) \\
0.296 \\ |0.0131p_1^{-1} +0.0261| \\ 0.0110 \\
0.0000850|\textrm{Im}(p_1)| \\
\end{smallmatrix}$ \\
\hline
\end{tabular}
\end{table}
\begin{table}[H]
\centering
\begin{tabular}{c|c|c||c|c|c} \hline
Label & \multirow{2}{*}{Observables} & \multirow{2}{*}{Values at $x_{bp}$} & Label & \multirow{2}{*}{Observables} & \multirow{2}{*}{Values at $x_{bp}$} \\
($x_{bp}$) & & & ($x_{bp}$) & & \\ \hline
{\scriptsize $\begin{array}{c}(2,8,\frac{1}{2})\bm{25}\\ (0.445)\\ \end{array}$} & $\begin{smallmatrix}
(2x^{15}, 2x^{7}) \\
(4x^{10}, x^{4}) \\
x^{2} \\ |p_1^{-1}x^{5} +2x^{5}| \\ x^{7} \\
2x^{14}|\textrm{Im}(p_1)| \\
\end{smallmatrix}$ & $\begin{smallmatrix}
( 1.06\times 10^{-5},  6.91\times 10^{-3}) \\
( 1.22\times 10^{-3},  3.92\times 10^{-2}) \\
0.198 \\ |0.0175p_1^{-1} +0.0349| \\ 0.00346 \\
0.0000239|\textrm{Im}(p_1)| \\
\end{smallmatrix}$ &
{\scriptsize $\begin{array}{c}(2,8,\frac{1}{2})\bm{26}\\ (0.445)\\ \end{array}$} & $\begin{smallmatrix}
(2x^{15}, 2x^{7}) \\
(2x^{8}, x^{4}) \\
x^{2} \\ |p_1^{-1}x^{5} +2x^{5}| \\ x^{7} \\
2x^{14}|\textrm{Im}(p_1)| \\
\end{smallmatrix}$ & $\begin{smallmatrix}
( 1.06\times 10^{-5},  6.91\times 10^{-3}) \\
( 3.08\times 10^{-3},  3.92\times 10^{-2}) \\
0.198 \\ |0.0175p_1^{-1} +0.0349| \\ 0.00346 \\
0.0000239|\textrm{Im}(p_1)| \\
\end{smallmatrix}$ \\
\hline
\end{tabular}
\end{table}

\begin{table}[H]
\caption{The leading terms of observables obtained from favorable structures in $(N_1,N_2,k)=(2,8,1)$.
The second and fifth columns show the structures of observables in the leading order approximation of $x\equiv|\varepsilon_1|=|\varepsilon_2|^k$.
The third and sixth columns show values of those at $x=x_{bp}$.
There are 10 different leading terms of observables labeled as $(2,8,1)\bm{n}$ $(\bm{n}=\bm{1},\bm{2},...,\bm{10})$.}
\label{tab:(2,8,1)}
\centering
\begin{tabular}{c|c|c||c|c|c} \hline
Label & \multirow{2}{*}{Observables} & \multirow{2}{*}{Values at $x_{bp}$} & Label & \multirow{2}{*}{Observables} & \multirow{2}{*}{Values at $x_{bp}$} \\
($x_{bp}$) & & & ($x_{bp}$) & & \\ \hline
{\scriptsize $\begin{array}{c}(2,8,1)\bm{1}\\ (0.220)\\ \end{array}$} & $\begin{smallmatrix}
(2x^{8}, x^{4}) \\
(4x^{6}, x^{2}) \\
4x^{2} \\  |2x^{3} +p_1^{-1}x^{3}| \\  |4x^{5} +4p_1x^{5}| \\
16x^{10}|\textrm{Im}(p_1)| \\
\end{smallmatrix}$ & $\begin{smallmatrix}
( 1.1\times 10^{-5},  2.34\times 10^{-3}) \\
( 4.54\times 10^{-4},  4.84\times 10^{-2}) \\
0.194 \\ |0.0213 +0.0106p_1^{-1}| \\ |0.00206 +0.00206p_1| \\
0.00000425|\textrm{Im}(p_1)| \\
\end{smallmatrix}$ &
{\scriptsize $\begin{array}{c}(2,8,1)\bm{2}\\ (0.220)\\ \end{array}$} & $\begin{smallmatrix}
(2x^{8}, x^{4}) \\
(2x^{5}, x^{2}) \\
4x^{2} \\  |2x^{3} +p_1^{-1}x^{3}| \\  |4x^{5} +4p_1x^{5}| \\
16x^{10}|\textrm{Im}(p_1)| \\
\end{smallmatrix}$ & $\begin{smallmatrix}
( 1.1\times 10^{-5},  2.34\times 10^{-3}) \\
( 1.03\times 10^{-3},  4.84\times 10^{-2}) \\
0.194 \\ |0.0213 +0.0106p_1^{-1}| \\ |0.00206 +0.00206p_1| \\
0.00000425|\textrm{Im}(p_1)| \\
\end{smallmatrix}$ \\
\hline
\end{tabular}
\end{table}

\begin{table}[H]
\caption{The leading terms of observables obtained from favorable structures in $(N_1,N_2,k)=(2,9,\frac{1}{2})$.
The second and fifth columns show the structures of observables in the leading order approximation of $x\equiv|\varepsilon_1|=|\varepsilon_2|^k$.
The third and sixth columns show values of those at $x=x_{bp}$.
There are 63 different leading terms of observables labeled as $(2,9,\frac{1}{2})\bm{n}$ $(\bm{n}=\bm{1},\bm{2},...,\bm{63})$.}
\label{tab:(2,9,1/2)}
\centering
\begin{tabular}{c|c|c||c|c|c} \hline
Label & \multirow{2}{*}{Observables} & \multirow{2}{*}{Values at $x_{bp}$} & Label & \multirow{2}{*}{Observables} & \multirow{2}{*}{Values at $x_{bp}$} \\
($x_{bp}$) & & & ($x_{bp}$) & & \\ \hline
{\scriptsize $\begin{array}{c}(2,9,\frac{1}{2})\bm{1}\\ (0.420)\\ \end{array}$} & $\begin{smallmatrix}
(4x^{16}, x^{8}) \\
(2x^{8}, x^{4}) \\
4x^{3} \\ |p_1^{-1}x^{5} +2x^{5}| \\  2x^{6} \\
16x^{14}|\textrm{Im}(p_1)| \\
\end{smallmatrix}$ & $\begin{smallmatrix}
( 3.75\times 10^{-6},  9.68\times 10^{-4}) \\
( 1.94\times 10^{-3},  3.11\times 10^{-2}) \\
0.296 \\ |0.0131p_1^{-1} +0.0261| \\ 0.0110 \\
0.0000850|\textrm{Im}(p_1)| \\
\end{smallmatrix}$ &
{\scriptsize $\begin{array}{c}(2,9,\frac{1}{2})\bm{2}\\ (0.420)\\ \end{array}$} & $\begin{smallmatrix}
(4x^{16}, x^{8}) \\
(4x^{10}, x^{4}) \\
4x^{3} \\ |p_1^{-1}x^{5} +2x^{5}| \\  2x^{6} \\
16x^{14}|\textrm{Im}(p_1)| \\
\end{smallmatrix}$ & $\begin{smallmatrix}
( 3.75\times 10^{-6},  9.68\times 10^{-4}) \\
( 6.83\times 10^{-4},  3.11\times 10^{-2}) \\
0.296 \\ |0.0131p_1^{-1} +0.0261| \\ 0.0110 \\
0.0000850|\textrm{Im}(p_1)| \\
\end{smallmatrix}$ \\
\hline
\end{tabular}
\end{table}
\begin{table}[H]
\centering
\begin{tabular}{c|c|c||c|c|c} \hline
Label & \multirow{2}{*}{Observables} & \multirow{2}{*}{Values at $x_{bp}$} & Label & \multirow{2}{*}{Observables} & \multirow{2}{*}{Values at $x_{bp}$} \\
($x_{bp}$) & & & ($x_{bp}$) & & \\ \hline
{\scriptsize $\begin{array}{c}(2,9,\frac{1}{2})\bm{5}\\ (0.420)\\ \end{array}$} & $\begin{smallmatrix}
(2x^{14}, x^{8}) \\
(4x^{10}, x^{4}) \\
 4x^{3} \\ |p_1^{-1}x^{5} +2x^{5}| \\ 2x^{6} \\
16x^{14}|\textrm{Im}(p_1)| \\
\end{smallmatrix}$ & $\begin{smallmatrix}
( 1.06\times 10^{-5},  9.68\times 10^{-4}) \\
( 6.83\times 10^{-4},  3.11\times 10^{-2}) \\
0.296 \\ |0.0131p_1^{-1} +0.0261| \\ 0.0110 \\
0.0000850|\textrm{Im}(p_1)| \\
\end{smallmatrix}$ &
{\scriptsize $\begin{array}{c}(2,9,\frac{1}{2})\bm{6}\\ (0.420)\\ \end{array}$} & $\begin{smallmatrix}
(2x^{14}, x^{8}) \\
(2x^{8}, x^{4}) \\
 4x^{3} \\ |p_1^{-1}x^{5} +2x^{5}| \\ 2x^{6} \\
16x^{14}|\textrm{Im}(p_1)| \\
\end{smallmatrix}$ & $\begin{smallmatrix}
( 1.06\times 10^{-5},  9.68\times 10^{-4}) \\
( 1.94\times 10^{-3},  3.11\times 10^{-2}) \\
0.296 \\ |0.0131p_1^{-1} +0.0261| \\ 0.0110 \\
0.0000850|\textrm{Im}(p_1)| \\
\end{smallmatrix}$ \\
\hline
\end{tabular}
\end{table}
\begin{table}[H]
\centering
\begin{tabular}{c|c|c||c|c|c} \hline
Label & \multirow{2}{*}{Observables} & \multirow{2}{*}{Values at $x_{bp}$} & Label & \multirow{2}{*}{Observables} & \multirow{2}{*}{Values at $x_{bp}$} \\
($x_{bp}$) & & & ($x_{bp}$) & & \\ \hline
{\scriptsize $\begin{array}{c}(2,9,\frac{1}{2})\bm{25}\\ (0.445)\\ \end{array}$} & $\begin{smallmatrix}
(2x^{15}, x^{8}) \\
(2x^{9}, x^{4}) \\
2x^{3} \\ |p_1^{-1}x^{5} +2x^{5}| \\ 2x^{8} \\
8x^{16}|\textrm{Im}(p_1)| \\
\end{smallmatrix}$ & $\begin{smallmatrix}
( 1.06\times 10^{-5},  1.54\times 10^{-3}) \\
( 1.37\times 10^{-3},  3.92\times 10^{-2}) \\
0.176 \\ |0.0175p_1^{-1} +0.0349| \\ 0.00308 \\
0.0000189|\textrm{Im}(p_1)| \\
\end{smallmatrix}$ &
{\scriptsize $\begin{array}{c}(2,9,\frac{1}{2})\bm{26}\\ (0.469)\\ \end{array}$} & $\begin{smallmatrix}
(2x^{16}, 2x^{9}) \\
(2x^{10}, x^{4}) \\
x^{3} \\ |p_1^{-1}x^{5} +2x^{5}| \\ x^{8} \\
x^{16}|\textrm{Im}(2p_1)| \\
\end{smallmatrix}$ & $\begin{smallmatrix}
( 1.1\times 10^{-5},  2.2\times 10^{-3}) \\
( 1.03\times 10^{-3},  4.84\times 10^{-2}) \\
0.103 \\ |0.0227p_1^{-1} +0.0454| \\ 0.00234 \\
0.0000110|\textrm{Im}(p_1)| \\
\end{smallmatrix}$ \\
\hline
\end{tabular}
\end{table}
\begin{table}[H]
\centering
\begin{tabular}{c|c|c||c|c|c} \hline
Label & \multirow{2}{*}{Observables} & \multirow{2}{*}{Values at $x_{bp}$} & Label & \multirow{2}{*}{Observables} & \multirow{2}{*}{Values at $x_{bp}$} \\
($x_{bp}$) & & & ($x_{bp}$) & & \\ \hline
{\scriptsize $\begin{array}{c}(2,9,\frac{1}{2})\bm{45}\\ (0.420)\\ \end{array}$} & $\begin{smallmatrix}
(x^{14}, 2x^{8}) \\
(x^{7}, 2x^{5}) \\
0.5x \\  2x^{4} \\ |p_1^{-1}x^{7} -2x^{7}| \\
2x^{12}|\textrm{Im}(p_1)| \\
\end{smallmatrix}$ & $\begin{smallmatrix}
( 5.31\times 10^{-6},  1.94\times 10^{-3}) \\
( 2.31\times 10^{-3},  2.61\times 10^{-2}) \\
0.210 \\ 0.0622 \\ |0.00231p_1^{-1} -0.00461| \\
0.0000603|\textrm{Im}(p_1^{-1})| \\
\end{smallmatrix}$ &
{\scriptsize $\begin{array}{c}(2,9,\frac{1}{2})\bm{46}\\ (0.420)\\ \end{array}$} & $\begin{smallmatrix}
(x^{14}, 2x^{8}) \\
(x^{8}, 2x^{5}) \\
0.5x \\  2x^{4} \\ |p_1^{-1}x^{7} -2x^{7}| \\
2x^{12}|\textrm{Im}(p_1)| \\
\end{smallmatrix}$ & $\begin{smallmatrix}
( 5.31\times 10^{-6},  1.94\times 10^{-3}) \\
( 9.68\times 10^{-4},  2.61\times 10^{-2}) \\
0.210 \\ 0.0622 \\ |0.00231p_1^{-1} -0.00461| \\
0.0000603|\textrm{Im}(p_1^{-1})| \\
\end{smallmatrix}$ \\
\hline
\end{tabular}
\end{table}

\begin{table}[H]
\caption{The leading terms of observables obtained from favorable structures in $(N_1,N_2,k)=(2,9,1)$.
The second and fifth columns show the structures of observables in the leading order approximation of $x\equiv|\varepsilon_1|=|\varepsilon_2|^k$.
The third and sixth columns show values of those at $x=x_{bp}$.
There are 19 different leading terms of observables labeled as $(2,9,1)\bm{n}$ $(\bm{n}=\bm{1},\bm{2},...,\bm{19})$.}
\label{tab:(2,9,1)}
\centering
\begin{tabular}{c|c|c||c|c|c} \hline
Label & \multirow{2}{*}{Observables} & \multirow{2}{*}{Values at $x_{bp}$} & Label & \multirow{2}{*}{Observables} & \multirow{2}{*}{Values at $x_{bp}$} \\
($x_{bp}$) & & & ($x_{bp}$) & & \\ \hline
{\scriptsize $\begin{array}{c}(2,9,1)\bm{1}\\ (0.220)\\ \end{array}$} & $\begin{smallmatrix}
(2x^{8}, x^{4}) \\
(4x^{6}, x^{2}) \\
4x^{2} \\ |2x^{3} +p_1^{-1}x^{3}| \\ |4x^{5} +4p_1x^{5}| \\
16x^{10}|\textrm{Im}(p_1)| \\
\end{smallmatrix}$ & $\begin{smallmatrix}
( 1.1\times 10^{-5},  2.34\times 10^{-3}) \\
( 4.54\times 10^{-4},  4.84\times 10^{-2}) \\
0.194 \\ |0.0213 +0.0106p_1^{-1}| \\ |0.00206 +0.00206p_1| \\
0.00000425|\textrm{Im}(p_1)| \\
\end{smallmatrix}$ &
{\scriptsize $\begin{array}{c}(2,9,1)\bm{2}\\ (0.220)\\ \end{array}$} & $\begin{smallmatrix}
(2x^{8}, x^{4}) \\
(2x^{5}, x^{2}) \\
4x^{2} \\ |2x^{3} +p_1^{-1}x^{3}| \\ |4x^{5} +4p_1x^{5}| \\
16x^{10}|\textrm{Im}(p_1)| \\
\end{smallmatrix}$ & $\begin{smallmatrix}
( 1.1\times 10^{-5},  2.34\times 10^{-3}) \\
( 1.03\times 10^{-3},  4.84\times 10^{-2}) \\
0.194 \\ |0.0213 +0.0106p_1^{-1}| \\ |0.00206 +0.00206p_1| \\
0.00000425|\textrm{Im}(p_1)| \\
\end{smallmatrix}$ \\
\hline
\end{tabular}
\end{table}
\begin{table}[H]
\centering
\begin{tabular}{c|c|c||c|c|c} \hline
Label & \multirow{2}{*}{Observables} & \multirow{2}{*}{Values at $x_{bp}$} & Label & \multirow{2}{*}{Observables} & \multirow{2}{*}{Values at $x_{bp}$} \\
($x_{bp}$) & & & ($x_{bp}$) & & \\ \hline
{\scriptsize $\begin{array}{c}(2,9,1)\bm{19}\\ (0.220)\\ \end{array}$} & $\begin{smallmatrix}
(x^{8}, 2x^{5}) \\
(2x^{5}, x^{2}) \\
 x \\  |2x^{3} +p_1^{-1}x^{3}| \\ |p_1^{-1}x^{4} -2x^{4}| \\
4x^{8}|\textrm{Im}(p_1)| \\
\end{smallmatrix}$ & $\begin{smallmatrix}
( 5.49\times 10^{-6},  1.03\times 10^{-3}) \\
( 1.03\times 10^{-3},  4.84\times 10^{-2}) \\
0.220 \\ |0.0213 +0.0106p_1^{-1}| \\ |0.00234p_1^{-1} -0.00469| \\
0.0000220|\textrm{Im}(p_1)| \\
\end{smallmatrix}$ & & \\ \hline
\end{tabular}
\end{table}

\begin{table}[H]
\centering
\begin{tabular}{c|c|c||c|c|c} \hline
Label & \multirow{2}{*}{Observables} & \multirow{2}{*}{Values at $x_{bp}$} & Label & \multirow{2}{*}{Observables} & \multirow{2}{*}{Values at $x_{bp}$} \\
($x_{bp}$) & & & ($x_{bp}$) & & \\ \hline
{\scriptsize $\begin{array}{c}(3,6,\frac{1}{2})\bm{1}\\ (0.364)\\ \end{array}$} & $\begin{smallmatrix}
(2x^{12}, x^{6}) \\
(2x^{7}, x^{3}) \\
x \\  2x^{4} \\ x^{6} \\
2x^{11}|\textrm{Im}(p_1)| \\
\end{smallmatrix}$ & $\begin{smallmatrix}
( 1.08\times 10^{-5},  2.33\times 10^{-3}) \\
( 1.69\times 10^{-3},  4.82\times 10^{-2}) \\
0.364 \\ 0.0351 \\ 0.00233 \\
0.0000297|\textrm{Im}(p_1)| \\
\end{smallmatrix}$ &
{\scriptsize $\begin{array}{c}(3,6,\frac{1}{2})\bm{2}\\ (0.364)\\ \end{array}$} & $\begin{smallmatrix}
(2x^{12}, x^{6}) \\
(4x^{9}, x^{3}) \\
x \\  2x^{4} \\ x^{6} \\
2x^{11}|\textrm{Im}(p_1)| \\
\end{smallmatrix}$ & $\begin{smallmatrix}
( 1.08\times 10^{-5},  2.33\times 10^{-3}) \\
( 4.49\times 10^{-4},  4.82\times 10^{-2}) \\
0.364 \\ 0.0351 \\ 0.00233 \\
0.0000297|\textrm{Im}(p_1)| \\
\end{smallmatrix}$ \\ \hline
{\scriptsize $\begin{array}{c}(3,6,\frac{1}{2})\bm{3}\\ (0.364)\\ \end{array}$} & $\begin{smallmatrix}
(2x^{12}, x^{6}) \\
(2x^{8}, x^{3}) \\
x \\  2x^{4} \\ x^{6} \\
2x^{11}|\textrm{Im}(p_1)| \\
\end{smallmatrix}$ & $\begin{smallmatrix}
( 1.08\times 10^{-5},  2.33\times 10^{-3}) \\
( 6.16\times 10^{-4},  4.82\times 10^{-2}) \\
0.364 \\ 0.0351 \\ 0.00233 \\
0.0000297|\textrm{Im}(p_1)| \\
\end{smallmatrix}$ & & \\ \hline
\end{tabular}
\caption{The leading terms of observables obtained from favorable structures in $(N_1,N_2,k)=(3,6,\frac{1}{2})$.
The second and fifth columns show the structures of observables in the leading order approximation of $x\equiv|\varepsilon_1|=|\varepsilon_2|^k$.
The third and sixth columns show values of those at $x=x_{bp}$.
There are three different leading terms of observables labeled as $(3,6,\frac{1}{2})\bm{n}$ $(\bm{n}=\bm{1},\bm{2},\bm{3})$.}
\label{tab:(3,6,1/2)}
\end{table}

\begin{table}[H]
\centering
\begin{tabular}{c|c|c||c|c|c} \hline
Label & \multirow{2}{*}{Observables} & \multirow{2}{*}{Values at $x_{bp}$} & Label & \multirow{2}{*}{Observables} & \multirow{2}{*}{Values at $x_{bp}$} \\
($x_{bp}$) & & & ($x_{bp}$) & & \\ \hline
{\scriptsize $\begin{array}{c}(3,6,1)\bm{1}\\ (0.0310)\\ \end{array}$} & $\begin{smallmatrix}
(2x^{3.5}, x^{2}) \\
(2x^{2}, \sqrt{2}x) \\
x^{0.5} \\ x \\  |-2x^{1.5} +p_1p_2^{-1}x^{1.5}| \\
2x^{3}|\textrm{Im}(p_1p_2^{-1})| \\
\end{smallmatrix}$ & $\begin{smallmatrix}
( 1.05\times 10^{-5},  9.61\times 10^{-4}) \\
( 1.92\times 10^{-3},  4.38\times 10^{-2}) \\
0.176 \\ 0.0310 \\ |-0.0109 +0.00546p_1p_2^{-1}| \\
0.0000596|\textrm{Im}(p_1p_2^{-1})| \\
\end{smallmatrix}$ &
{\scriptsize $\begin{array}{c}(3,6,1)\bm{2}\\ (0.0310)\\ \end{array}$} & $\begin{smallmatrix}
(2x^{3.5}, x^{2}) \\
(2x^{2}, \sqrt{2}x) \\
x^{0.5} \\ x \\  |-2x^{1.5} +p_1^{-1}p_2x^{1.5}| \\
2x^{3}|\textrm{Im}(p_1p_2^{-1})| \\
\end{smallmatrix}$ & $\begin{smallmatrix}
( 1.05\times 10^{-5},  9.61\times 10^{-4}) \\
( 1.92\times 10^{-3},  4.38\times 10^{-2}) \\
0.176 \\ 0.0310 \\ |-0.0109 +0.00546p_1^{-1}p_2| \\
0.0000596|\textrm{Im}(p_1p_2^{-1})| \\
\end{smallmatrix}$ \\ \hline
\end{tabular}
\caption{The leading terms of observables obtained from favorable structures in $(N_1,N_2,k)=(3,6,2)$.
The second and fifth columns show the structures of observables in the leading order approximation of $x\equiv|\varepsilon_1|=|\varepsilon_2|^k$.
The third and sixth columns show values of those at $x=x_{bp}$.
There are two different leading terms of observables labeled as $(3,6,2)\bm{n}$ $(\bm{n}=\bm{1},\bm{2})$.}
\label{tab:(3,6,2)}
\end{table}


\subsection{The origins of the large mass hierarchy and CP violation}

Next we discuss the origins of the large mass hierarchy and CP violation in favorable structures of quark mass matrices.
As we have seen in previous subsection, favorable structures are obtained in seven types $(N_1,N_2,k)=(2,7,\frac{1}{2})$, $(2,8,\frac{1}{2})$, $(2,8,1)$, $(2,9,\frac{1}{2})$, $(2,9,1)$, $(3,6,\frac{1}{2})$ and $(3,6,2)$.
We discuss what is important in the large mass hierarchy and CP violation.
The orders of $x_{bp}$, relations between $|\varepsilon_1|$ and $|\varepsilon_2|$, and contributions of $p_1$ and $p_2$ to the leading terms of $J_\textrm{CP}$ in favorable structures may be useful.
In Table \ref{tab:size and p1 p2 dependency}, we summarize them in each type.

\begin{table}[H]
\centering
\begin{tabular}{c|ccccccc} \hline
$(N_1,N_2,k)$ & $(2,7,\frac{1}{2})$ & $(2,8,\frac{1}{2})$ & $(2,8,1)$ & $(2,9,\frac{1}{2})$ & $(2,9,1)$ & $(3,6,\frac{1}{2})$ & $(3,6,2)$ \\ \hline
${\cal O}(x_{bp})$ & 0.1 & 0.1 & 0.1 & 0.1 & 0.1 & 0.1 & 0.01 \\
$|\varepsilon_1|$ vs. $|\varepsilon_2|$ & $|\varepsilon_1|>|\varepsilon_2|$ & $|\varepsilon_1|>|\varepsilon_2|$ & $|\varepsilon_1|=|\varepsilon_2|$ & $|\varepsilon_1|>|\varepsilon_2|$ & $|\varepsilon_1|=|\varepsilon_2|$ & $|\varepsilon_1|>|\varepsilon_2|$ & $|\varepsilon_1|<|\varepsilon_2|$ \\
$p_1$, $p_2$ to $J_\textrm{CP}$ & $\textrm{Im}(p_1)$ & $\textrm{Im}(p_1)$ & $\textrm{Im}(p_1)$ & $\textrm{Im}(p_1)$ & $\textrm{Im}(p_1)$ & $\textrm{Im}(p_1)$ & $\textrm{Im}(p_1p_2^{-1})$ \\ \hline
\end{tabular}
\caption{The orders of $x_{bp}$, relations between $|\varepsilon_1|$ and $|\varepsilon_2|$, and contributions of $p_1$ and $p_2$ to the leading terms of $J_\textrm{CP}$ in favorable structures.
First row shows the orders of $x_{bp}$.
Second row shows size relation between $|\varepsilon_1|$ and $|\varepsilon_2|$.
Third row shows contributions of $p_1$ and $p_2$ to $J_\textrm{CP}$.}
\label{tab:size and p1 p2 dependency}
\end{table}
Table \ref{tab:size and p1 p2 dependency} implies that
\begin{itemize}
\item quark mass hierarchies mainly originate from $|\varepsilon_2|$ ($T_2$ ($Z_{N_2}$)-symmetry) since $|\varepsilon_1|\geq |\varepsilon_2|$ and $N_1 < N_2$;
\item CP violation originates from $p_1$ ($T_1$ ($Z_{N_1}$)-symmetry) since the leading terms of $J_\textrm{CP}$ do not depend on $p_2$;
\item suitable order of $|\varepsilon_1|$ is $0.1$,
\end{itemize}
in six types $(N_1,N_2,k)=(2,7,\frac{1}{2})$, $(2,8,\frac{1}{2})$, $(2,8,1)$, $(2,9,\frac{1}{2})$, $(2,9,1)$ and $(3,6,\frac{1}{2})$.
Note that mass matrix elements can contain $|\varepsilon_1|^r$ ($r=0,1,...,N_1-1$) and $|\varepsilon_2|^s$ ($s=0,1,...,N_2-1$) depending on their $(T_1,T_2)$-charges.
Therefore when $|\varepsilon_1|\geq |\varepsilon_2|$ and $N_1 < N_2$, it is plausible to regard the origin of mass hierarchies as $|\varepsilon_2|$.

In our analysis, six types have this kind of origins.
However, type $(N_1,N_2>N_1,k\leq 1)$ would give favorable structures which have same kind of origins.

Meanwhile, the origins of the large mass hierarchy and CP violation in favorable structures in type $(N_1,N_2,k)=(3,6,2)$ seem to be different from other six types.
To clarify it, let us consider Yukawa coupling with $(T_1,T_2)$-charges $(m,n)$ in type $(N_1,N_2,k)=(3,6,2)$.
It is deformed as
\begin{align}
&\begin{aligned}
Y(\tau_1,\tau_2) &\sim \varepsilon_1^m \cdot \varepsilon_2^n \\
&= e^{2\pi im\tau_1/3} \cdot e^{2\pi in\tau_2/6} \\
&= e^{2\pi im(\tau_1-\tau_2)/3} \cdot e^{2\pi i(n+2m)\tau_2/6} \\
&= {\varepsilon_1'}^m \cdot {\varepsilon_2'}^{n+2m} \\
&= (p_1p_2^{-1})^{m/3} |\varepsilon_1'|^m \cdot p_2^{(n+2m)/6} |\varepsilon_2'|^{n+2m}, \\
&(\varepsilon_1' \equiv \varepsilon_1/\varepsilon_2 = e^{2\pi i(\tau_1-\tau_2)/3}, 
\quad \varepsilon_2'\equiv \varepsilon_2 = e^{2\pi i\tau_2/6}), \label{eq:deformation}
\end{aligned}
\end{align}
where $m\in Z_{3}$ and $n\in Z_{6}$.
Note that $|\varepsilon_1'|=|\varepsilon_2'|$ since $|\varepsilon_1|=|\varepsilon_2|^2$.
Defining $T_1'\equiv T_1$, $T_2'\equiv T_1T_2$, $\tau_1'\equiv\tau_1-\tau_2$ and $\tau_2'\equiv\tau_2$, we obtain
\begin{align}
&T_1':(\tau_1',\tau_2') \to (\tau_1'+1,\tau_2'), \quad
T_2':(\tau_1',\tau_2') \to (\tau_1',\tau_2'+1), \\
&(\varepsilon_1', \varepsilon_2') \xrightarrow{T_1'} (e^{2\pi i/3}\varepsilon_1', \varepsilon_2'), \quad
(\varepsilon_1', \varepsilon_2') \xrightarrow{T_2'} (\varepsilon_1', e^{2\pi i/6}\varepsilon_2').
\end{align}
Then $Y(\tau_1,\tau_2)={\varepsilon_1'}^m \cdot {\varepsilon_2'}^{n+2m}$ has charges $m$ and $n+2m$ for $T_1'$ ($Z_3$) and $T_2'$ ($Z_6$)-symmetries, respectively.
At first sight, it seems that the deformation in Eq.~(\ref{eq:deformation}) means that Yukawa coupling with $(T_1,T_2)$-charges $(m,n)$ in type $(N_1,N_2,k)=(3,6,2)$ is equivalent to one with $(T_1',T_2')$-charges $(m,n+2m)$ in type $(N_1,N_2,k)=(3,6,1)$.
Nevertheless, we cannot find favorable structures in type $(N_1,N_2,k)=(3,6,1)$ as shown in Table \ref{tab:number of models leading to J_CP}.
This is because the powers of ${\varepsilon_2}'$, $n+2m$, run $\{0,1,...,9\}$ but ones of $\varepsilon_2$ in $(N_1,N_2,k)=(3,6,1)$ run $\{0,1,...,5\}$.
The powers in the former are enhanced by $T_1$-charge $m$ beyond $n\in Z_6$.
Obviously it is more difficult to realize the large mass hierarchy in the latter.
Actually, favorable structures exist in $(N_1,N_2,k)=(3,6,2)$ but do not exist in $(3,6,1)$.
In this sense, favorable structures in both types are not equivalent.

Now, we can estimate the origins of the large mass hierarchy and CP violation in favorable structures in type $(N_1,N_2,k)=(3,6,2)$.
Table \ref{tab:size and p1 p2 dependency} and discussion above imply that
\begin{itemize}
\item quark mass hierarchies mainly originate from $|\varepsilon_2'|$ ($T_2'$ ($Z_{6}$)-symmetry) since $|\varepsilon_1'| = |\varepsilon_2'|$ and $N_1=3 < N_2=6$.
Then the powers of $|\varepsilon_2'|$ are enhanced by $T_1$-symmetry as $|\varepsilon_2'|^{n+2m}$;
\item CP violation originates from $p_1p_2^{-1}$ ($T_1'$ ($Z_{3}$)-symmetry) since the leading terms of $J_\textrm{CP}$ depend on $\textrm{Im}(p_1p_2^{-1})$;
\item suitable order of $|\varepsilon_1'|=|\varepsilon_2| =|\varepsilon_1|^{1/2}$ is $0.1$,
\end{itemize}
in type $(N_1,N_2,k)=(3,6,2)$.

In our analysis, only $(N_1,N_2,k)=(3,6,2)$ has this kind of origins.
However, there may exist further possibilities possessing such origins.
Let us consider Yukawa coupling with $(T_1,T_2)$-charge $(m,n)$ in type $(N_1,N_2=gN_1,k>1)$, where $g\in \{2,3,...\}$.
It is deformed as
\begin{align}
\begin{aligned}
Y(\tau_1,\tau_2) &\sim \varepsilon_1^m \cdot \varepsilon_2^n \\
&=e^{2\pi im\tau_1/N_1} \cdot e^{2\pi in\tau_2/N_2} \\
&=e^{2\pi im(\tau_1-\ell\tau_2)/N_1} \cdot e^{2\pi i(n+m\ell g)\tau_2/N_2} \\
&= {\varepsilon_1'}^m \cdot {\varepsilon_2'}^{n+m\ell g} \\
&= (p_1p_2^{-\ell})^{m/N_1} |\varepsilon_1'|^m \cdot p_2^{(n+m\ell g)/N_2} |\varepsilon_2'|^{n+m\ell g}, \\
&(\varepsilon_1' \equiv \varepsilon_1/\varepsilon_2^\ell, \quad \varepsilon_2' \equiv \varepsilon_2, \quad k > \ell, \quad \ell \in \mathbb{Z}_+),
\end{aligned}
\end{align}
where $m\in Z_{N_1}$ and $n\in Z_{N_2}$.
Note that $|\varepsilon_1'|=|\varepsilon_2'|^{k'}$, $k'=k-\ell$ since $|\varepsilon_1|=|\varepsilon_2|^k$.
Defining $T_1'\equiv T_1$, $T_2'\equiv T_1^\ell T_2$, $\tau_1'\equiv\tau_1-\ell\tau_2$ and $\tau_2'\equiv\tau_2$, we obtain
\begin{align}
&T_1':(\tau_1',\tau_2') \to (\tau_1'+1,\tau_2'), \quad
T_2':(\tau_1',\tau_2') \to (\tau_1',\tau_2'+1), \\
&(\varepsilon_1', \varepsilon_2') \xrightarrow{T_1'} (e^{2\pi i/N_1}\varepsilon_1', \varepsilon_2'), \quad
(\varepsilon_1', \varepsilon_2') \xrightarrow{T_2'} (\varepsilon_1', e^{2\pi i/N_2}\varepsilon_2').
\end{align}
Then $Y(\tau_1,\tau_2)={\varepsilon_1'}^m \cdot {\varepsilon_2'}^{n+m\ell g}$ has charges $m$ and $n+m\ell g$ for $T_1'$ ($Z_{N_1}$) and $T_2'$ ($Z_{N_2}$)-symmetries, respectively.
Thus, type $(N_1,N_2=gN_1,k>1)$ would give favorable structures which have the following origins,
\begin{itemize}
\item quark mass hierarchies mainly originate from $|\varepsilon_2'|$ ($T_2'$ ($Z_{N_1}$)-symmetry).
Then the powers of $|\varepsilon_2'|$ are enhanced by $T_1$-symmetry as $|\varepsilon_2'|^{n+m\ell g}$;
\item CP violation originates from $p_1p_2^{-\ell}$ ($T_1'$ ($Z_{N_1}$)-symmetry);
\item suitable order of $|\varepsilon_1'|=|\varepsilon_2| =|\varepsilon_1|^{1/k'}$ is $x_{bp}^{1/k'}$.
\end{itemize}


\section{Conclusion}
\label{sec:conclusion}

We have studied the quark mass matrices with two moduli near the modular fixed points.
Multi-moduli are required from the viewpoint of sizable CP violation.
We have concentrated on the vicinity of the cusp, $(\tau_1,\tau_2)\sim(i\infty,i\infty)$, where residual $T_1$ $(Z_{N_1})$ and $T_2$ $(Z_{N_2})$-symmetries remains.
Then mass matrix elements are approximately given by $\varepsilon_1^{m}\cdot \varepsilon_2^{n}$, where $m$ and $n$ are their residual $(T_1,T_2)$-charges.
Similar structures of mass matrices are also obtained in $Z_{N_1}\times Z_{N_2}$ Froggatt-Nielsen models.
A smaller residual symmetry $Z_{N_1}$ with $N_1=2,3,4$ can also originate from $S_1$ and $S_1T_1$ symmetries at 
$\tau_1 = i $ and $e^{2\pi i/3}$.
A larger residual symmetry $Z_{N_2}$ may originate from a proper combination of $Z_{N_3} \times Z_{N_4}$, where $N_3,N_4=2,3,4$.

To find the phenomenologically favorable structures of mass matrices with two moduli, we have explored possible patterns of $(T_1,T_2)$-charges of $Q_i$, $u_i$ and $d_i$.
Our purpose is to reproduce both large mass hierarchy and CP violation without fine-tuning.
For our purpose, we have restricted the values of coupling coefficients in mass matrices to $\pm 1$.
In addition, we have introduced four conditions: the conditions for hierarchical masses in Eq.~(\ref{eq:conditions_of_powers}), the necessary conditions for CP violation in Eqs.~(\ref{eq:CP1}), (\ref{eq:CP2}), (\ref{eq:CP3}) and (\ref{eq:CP4}), the conditions to realize the orders of the quark mass ratios and the absolute values of the CKM matrix elements in Eq.~(\ref{eq:conditions_mass_CKM}), and the condition to realize the order of Jarlskog invariant.
As a result, we have found the phenomenologically favorable structures of mass matrices in seven types $(N_1,N_2,k)=(2,7,\frac{1}{2})$, $(2,8,\frac{1}{2})$, $(2,8,1)$, $(2,9,\frac{1}{2})$, $(2,9,1)$, $(3,6,\frac{1}{2})$ and $(3,6,2)$, where $k$ is defined by the relation $|\varepsilon_1|=|\varepsilon_2|^k$.
They have non-vanishing Jarlskog invariant $J_\textrm{CP}$ in the leading order approximation of $x\equiv |\varepsilon_1|=|\varepsilon_2|^k$.
Therefore, sizable CP violation can be expected.

Finally we have discussed the origins of the large mass hierarchy and CP violation in favorable structures of quark mass matrices.
There are two kinds of origins.
In types $(N_1,N_2>N_1,k\leq 1)$, the large quark mass hierarchy mainly originates from $T_2$-symmetry while CP violation is induced by $T_1$-symmetry.
In another types $(N_1,N_2=gN_1,k>1)$ with $g\in\{2,3,...\}$, the large quark mass hierarchy mainly originates from $T_2'$-symmetry while CP violation is induced by $T_1'$-symmetry.
Then mass hierarchy is also enhanced by $T_1$-symmetry.

Again we note that our analysis can be applied for $Z_{N_1}\times Z_{N_2}$ Froggatt-Nielsen models.
Thus we can find same favorable structures and origins of large mass hierarchy and CP violation in Froggatt-Nielsen models.
Using our results, promising models can be found in both modular symmetric flavor models and Froggatt-Nielsen models with $Z_{N_1}\times Z_{N_2}$-symmetry.
We leave the studies in explicit models of those for future study.


\vspace{1.5 cm}
\noindent
{\large\bf Acknowledgement}\\

This work was supported by JSPS KAKENHI Grant Numbers JP22KJ0047 (S. K.) and JP23K03375 (T. K.), and JST SPRING Grant Number JPMJSP2119(K. N.).

\appendix
\section*{Appendix}


\section{Patterns of $(T_1,T_2)$-charges and signs in coupling coefficients giving favorable structures}
\label{app:patterns of charges}

Here we show patterns of $(T_1,T_2)$-charges and signs in coupling coefficients giving favorable structures.
We show ones of $(N_1,N_2,k)=(2,7,\frac{1}{2})$ in Table \ref{tab:T-charges (2,7,1/2)}, $(2,8,\frac{1}{2})$ in Table \ref{tab:T-charges (2,8,1/2)}, $(2,8,1)$ in Table \ref{tab:T-charges (2,8,1)}, $(2,9,\frac{1}{2})$ in Table \ref{tab:T-charges (2,9,1/2)}, $(2,9,1)$ in Table \ref{tab:T-charges (2,9,1)}, $(3,6,\frac{1}{2})$ in Table \ref{tab:T-charges (3,6,1/2)} and $(3,6,2)$ in Table \ref{tab:T-charges (3,6,2)}, respectively.

\begin{table}[H]
\caption{$(T_1,T_2)$-charges and coupling coefficients in type $(N_1,N_2,k)=(2,7,\frac{1}{2})$ giving favorable structures.
The last column shows the favorable structures given in each pattern of $(T_1,T_2)$-charges and coupling coefficients.}
\label{tab:T-charges (2,7,1/2)}
\centering
\footnotesize
\begin{tabularx}{180mm}{CCCCCCCCCCCCCCCCCCCCCC|c} \hline
$p_{Q11}$ & $p_{Q21}$ & $p_{Q12}$ & $p_{Q22}$ & $p_{u11}$ & $p_{u21}$ & $p_{u12}$ & $p_{u22}$ & $p_{d11}$ & $p_{d21}$ & $p_{d12}$ & $p_{d22}$ & $a_{22}$ & $a_{23}$ & $a_{32}$ & $a_{33}$ & $b_{21}$ & $b_{22}$ & $b_{23}$ & $b_{31}$ & $b_{32}$ & $b_{33}$ & Label \\ \hline
1 & 0 & 3 & 2 & 0 & 0 & 3 & 2 & 1 & 1 & 1 & 0 & -1 & 1 & -1 & -1 & 1 & -1 & 1 & 1 & 1 & 1 & $(2,7,\frac{1}{2})\bm{1}$ \\
1 & 0 & 3 & 2 & 0 & 0 & 3 & 2 & 1 & 1 & 1 & 0 & -1 & 1 & -1 & -1 & -1 & -1 & 1 & -1 & 1 & 1 & $(2,7,\frac{1}{2})\bm{1}$ \\
1 & 0 & 3 & 2 & 0 & 0 & 3 & 2 & 1 & 1 & 1 & 0 & 1 & -1 & -1 & -1 & -1 & 1 & -1 & 1 & 1 & 1 & $(2,7,\frac{1}{2})\bm{1}$ \\
1 & 0 & 3 & 2 & 0 & 0 & 3 & 2 & 1 & 1 & 1 & 0 & 1 & -1 & -1 & -1 & 1 & 1 & -1 & -1 & 1 & 1 & $(2,7,\frac{1}{2})\bm{1}$ \\
1 & 0 & 3 & 2 & 0 & 0 & 3 & 2 & 1 & 1 & 1 & 0 & -1 & 1 & -1 & -1 & 1 & -1 & 1 & -1 & 1 & 1 & $(2,7,\frac{1}{2})\bm{2}$ \\
1 & 0 & 3 & 2 & 0 & 0 & 3 & 2 & 1 & 1 & 1 & 0 & 1 & -1 & -1 & -1 & -1 & 1 & -1 & -1 & 1 & 1 & $(2,7,\frac{1}{2})\bm{2}$ \\
\hline
\end{tabularx}
\end{table}

\begin{table}[H]
\caption{$(T_1,T_2)$-charges and coupling coefficients in type $(N_1,N_2,k)=(2,8,\frac{1}{2})$ giving favorable structures.
The last column shows the favorable structures given in each pattern of $(T_1,T_2)$-charges and coupling coefficients.}
\label{tab:T-charges (2,8,1/2)}
\centering
\footnotesize
\begin{tabularx}{180mm}{CCCCCCCCCCCCCCCCCCCCCC|c} \hline
$p_{Q11}$ & $p_{Q21}$ & $p_{Q12}$ & $p_{Q22}$ & $p_{u11}$ & $p_{u21}$ & $p_{u12}$ & $p_{u22}$ & $p_{d11}$ & $p_{d21}$ & $p_{d12}$ & $p_{d22}$ & $a_{22}$ & $a_{23}$ & $a_{32}$ & $a_{33}$ & $b_{21}$ & $b_{22}$ & $b_{23}$ & $b_{31}$ & $b_{32}$ & $b_{33}$ & Label \\ \hline
0 & 1 & 3 & 2 & 0 & 1 & 4 & 2 & 0 & 1 & 1 & 0 & -1 & -1 & -1 & 1 & 1 & 1 & -1 & 1 & 1 & -1 & $(2,8,\frac{1}{2})\bm{1}$ \\
0 & 1 & 3 & 2 & 0 & 1 & 4 & 2 & 0 & 1 & 1 & 0 & -1 & -1 & -1 & 1 & -1 & 1 & -1 & 1 & 1 & -1 & $(2,8,\frac{1}{2})\bm{1}$ \\
0 & 1 & 3 & 2 & 0 & 1 & 4 & 2 & 0 & 1 & 1 & 0 & -1 & -1 & -1 & 1 & 1 & 1 & -1 & -1 & 1 & -1 & $(2,8,\frac{1}{2})\bm{2}$ \\
0 & 1 & 3 & 2 & 0 & 1 & 4 & 2 & 1 & 1 & 1 & 0 & -1 & -1 & -1 & 1 & -1 & 1 & -1 & 1 & 1 & -1 & $(2,8,\frac{1}{2})\bm{3}$ \\
0 & 1 & 3 & 2 & 0 & 1 & 4 & 2 & 1 & 1 & 1 & 0 & -1 & -1 & -1 & 1 & 1 & 1 & -1 & -1 & 1 & -1 & $(2,8,\frac{1}{2})\bm{4}$ \\
0 & 1 & 3 & 2 & 0 & 1 & 4 & 2 & 1 & 1 & 1 & 0 & -1 & -1 & -1 & 1 & -1 & 1 & -1 & -1 & 1 & -1 & $(2,8,\frac{1}{2})\bm{4}$ \\
0 & 1 & 3 & 2 & 0 & 1 & 4 & 2 & 0 & 1 & 1 & 0 & 1 & 1 & 1 & -1 & -1 & -1 & 1 & 1 & -1 & 1 & $(2,8,\frac{1}{2})\bm{5}$ \\
\hline
\end{tabularx}
\end{table}
\begin{table}[H]
\centering
\footnotesize
\begin{tabularx}{180mm}{CCCCCCCCCCCCCCCCCCCCCC|c} \hline
$p_{Q11}$ & $p_{Q21}$ & $p_{Q12}$ & $p_{Q22}$ & $p_{u11}$ & $p_{u21}$ & $p_{u12}$ & $p_{u22}$ & $p_{d11}$ & $p_{d21}$ & $p_{d12}$ & $p_{d22}$ & $a_{22}$ & $a_{23}$ & $a_{32}$ & $a_{33}$ & $b_{21}$ & $b_{22}$ & $b_{23}$ & $b_{31}$ & $b_{32}$ & $b_{33}$ & Label \\ \hline
0 & 1 & 3 & 2 & 1 & 1 & 4 & 2 & 0 & 1 & 1 & 0 & 1 & -1 & -1 & -1 & -1 & -1 & 1 & -1 & 1 & -1 & $(2,8,\frac{1}{2})\bm{16}$ \\
0 & 1 & 3 & 2 & 1 & 1 & 4 & 2 & 0 & 1 & 1 & 0 & -1 & -1 & -1 & -1 & -1 & -1 & 1 & -1 & 1 & -1 & $(2,8,\frac{1}{2})\bm{16}$ \\
0 & 1 & 3 & 2 & 1 & 1 & 4 & 2 & 0 & 1 & 2 & 0 & 1 & 1 & -1 & -1 & 1 & 1 & -1 & 1 & 1 & -1 & $(2,8,\frac{1}{2})\bm{17}$ \\
0 & 1 & 3 & 2 & 1 & 1 & 4 & 2 & 0 & 1 & 2 & 0 & 1 & 1 & -1 & -1 & -1 & 1 & -1 & 1 & 1 & -1 & $(2,8,\frac{1}{2})\bm{17}$ \\
0 & 1 & 3 & 2 & 1 & 1 & 4 & 2 & 0 & 1 & 2 & 0 & -1 & 1 & -1 & -1 & 1 & 1 & -1 & 1 & 1 & -1 & $(2,8,\frac{1}{2})\bm{17}$ \\
0 & 1 & 3 & 2 & 1 & 1 & 4 & 2 & 0 & 1 & 2 & 0 & -1 & 1 & -1 & -1 & -1 & 1 & -1 & 1 & 1 & -1 & $(2,8,\frac{1}{2})\bm{17}$ \\
0 & 1 & 3 & 2 & 1 & 1 & 4 & 2 & 0 & 1 & 2 & 0 & 1 & -1 & -1 & -1 & 1 & -1 & 1 & 1 & 1 & -1 & $(2,8,\frac{1}{2})\bm{17}$ \\
\hline
\end{tabularx}
\end{table}
\begin{table}[H]
\centering
\footnotesize
\begin{tabularx}{180mm}{CCCCCCCCCCCCCCCCCCCCCC|c} \hline
$p_{Q11}$ & $p_{Q21}$ & $p_{Q12}$ & $p_{Q22}$ & $p_{u11}$ & $p_{u21}$ & $p_{u12}$ & $p_{u22}$ & $p_{d11}$ & $p_{d21}$ & $p_{d12}$ & $p_{d22}$ & $a_{22}$ & $a_{23}$ & $a_{32}$ & $a_{33}$ & $b_{21}$ & $b_{22}$ & $b_{23}$ & $b_{31}$ & $b_{32}$ & $b_{33}$ & Label \\ \hline
1 & 1 & 3 & 2 & 0 & 0 & 4 & 1 & 1 & 1 & 2 & 0 & 1 & -1 & -1 & -1 & 1 & -1 & 1 & 1 & 1 & -1 & $(2,8,\frac{1}{2})\bm{28}$ \\
1 & 1 & 3 & 2 & 0 & 0 & 4 & 1 & 1 & 1 & 2 & 0 & 1 & -1 & -1 & -1 & 1 & -1 & 1 & -1 & 1 & -1 & $(2,8,\frac{1}{2})\bm{28}$ \\
1 & 1 & 3 & 2 & 0 & 0 & 4 & 2 & 0 & 1 & 1 & 0 & -1 & 1 & -1 & -1 & -1 & -1 & 1 & 1 & -1 & 1 & $(2,8,\frac{1}{2})\bm{29}$ \\
1 & 1 & 3 & 2 & 0 & 0 & 4 & 2 & 0 & 1 & 1 & 0 & -1 & 1 & -1 & -1 & -1 & -1 & 1 & -1 & -1 & 1 & $(2,8,\frac{1}{2})\bm{29}$ \\
1 & 1 & 3 & 2 & 0 & 0 & 4 & 2 & 0 & 1 & 1 & 0 & 1 & -1 & -1 & -1 & 1 & 1 & -1 & 1 & -1 & 1 & $(2,8,\frac{1}{2})\bm{29}$ \\
1 & 1 & 3 & 2 & 0 & 0 & 4 & 2 & 0 & 1 & 1 & 0 & 1 & -1 & -1 & -1 & 1 & 1 & -1 & -1 & -1 & 1 & $(2,8,\frac{1}{2})\bm{29}$ \\
1 & 1 & 3 & 2 & 0 & 0 & 4 & 2 & 0 & 1 & 1 & 0 & -1 & 1 & -1 & -1 & 1 & 1 & -1 & 1 & 1 & -1 & $(2,8,\frac{1}{2})\bm{30}$ \\
\hline
\end{tabularx}
\end{table}
\begin{table}[H]
\centering
\footnotesize
\begin{tabularx}{180mm}{CCCCCCCCCCCCCCCCCCCCCC|c} \hline
$p_{Q11}$ & $p_{Q21}$ & $p_{Q12}$ & $p_{Q22}$ & $p_{u11}$ & $p_{u21}$ & $p_{u12}$ & $p_{u22}$ & $p_{d11}$ & $p_{d21}$ & $p_{d12}$ & $p_{d22}$ & $a_{22}$ & $a_{23}$ & $a_{32}$ & $a_{33}$ & $b_{21}$ & $b_{22}$ & $b_{23}$ & $b_{31}$ & $b_{32}$ & $b_{33}$ & Label \\ \hline
1 & 1 & 3 & 2 & 1 & 0 & 4 & 1 & 0 & 1 & 1 & 0 & -1 & 1 & -1 & -1 & -1 & -1 & 1 & 1 & -1 & 1 & $(2,8,\frac{1}{2})\bm{39}$ \\
1 & 1 & 3 & 2 & 1 & 0 & 4 & 1 & 0 & 1 & 1 & 0 & 1 & -1 & -1 & -1 & 1 & 1 & -1 & 1 & -1 & 1 & $(2,8,\frac{1}{2})\bm{39}$ \\
1 & 1 & 3 & 2 & 1 & 0 & 4 & 1 & 1 & 1 & 1 & 0 & -1 & 1 & 1 & 1 & -1 & 1 & -1 & -1 & -1 & 1 & $(2,8,\frac{1}{2})\bm{40}$ \\
1 & 1 & 3 & 2 & 1 & 0 & 4 & 1 & 1 & 1 & 1 & 0 & 1 & -1 & 1 & 1 & 1 & -1 & 1 & -1 & -1 & 1 & $(2,8,\frac{1}{2})\bm{40}$ \\
1 & 1 & 3 & 2 & 1 & 0 & 4 & 1 & 1 & 1 & 1 & 0 & -1 & 1 & -1 & -1 & -1 & 1 & -1 & 1 & 1 & -1 & $(2,8,\frac{1}{2})\bm{40}$ \\
1 & 1 & 3 & 2 & 1 & 0 & 4 & 1 & 1 & 1 & 1 & 0 & 1 & -1 & -1 & -1 & 1 & -1 & 1 & 1 & 1 & -1 & $(2,8,\frac{1}{2})\bm{40}$ \\
1 & 1 & 3 & 2 & 1 & 0 & 4 & 1 & 1 & 1 & 1 & 0 & -1 & 1 & 1 & 1 & 1 & -1 & 1 & -1 & 1 & -1 & $(2,8,\frac{1}{2})\bm{41}$ \\
\hline
\end{tabularx}
\end{table}

\begin{table}[H]
\caption{$(T_1,T_2)$-charges and coupling coefficients in type $(N_1,N_2,k)=(2,8,1)$ giving favorable structures.
The last column shows the favorable structures given in each pattern of $(T_1,T_2)$-charges and coupling coefficients.}
\label{tab:T-charges (2,8,1)}
\centering
\footnotesize
\begin{tabularx}{180mm}{CCCCCCCCCCCCCCCCCCCCCC|c} \hline
$p_{Q11}$ & $p_{Q21}$ & $p_{Q12}$ & $p_{Q22}$ & $p_{u11}$ & $p_{u21}$ & $p_{u12}$ & $p_{u22}$ & $p_{d11}$ & $p_{d21}$ & $p_{d12}$ & $p_{d22}$ & $a_{22}$ & $a_{23}$ & $a_{32}$ & $a_{33}$ & $b_{21}$ & $b_{22}$ & $b_{23}$ & $b_{31}$ & $b_{32}$ & $b_{33}$ & Label \\ \hline
0 & 1 & 3 & 2 & 1 & 1 & 4 & 2 & 0 & 1 & 1 & 0 & -1 & 1 & -1 & 1 & 1 & 1 & -1 & 1 & -1 & 1 & $(2,8,1)\bm{1}$ \\
0 & 1 & 3 & 2 & 1 & 1 & 4 & 2 & 0 & 1 & 1 & 0 & 1 & -1 & -1 & 1 & -1 & -1 & 1 & 1 & -1 & 1 & $(2,8,1)\bm{1}$ \\
0 & 1 & 3 & 2 & 1 & 1 & 4 & 2 & 0 & 1 & 2 & 0 & -1 & 1 & -1 & 1 & 1 & 1 & -1 & -1 & -1 & 1 & $(2,8,1)\bm{2}$ \\
0 & 1 & 3 & 2 & 1 & 1 & 4 & 2 & 0 & 1 & 2 & 0 & -1 & 1 & -1 & 1 & -1 & 1 & -1 & -1 & -1 & 1 & $(2,8,1)\bm{2}$ \\
0 & 1 & 3 & 2 & 1 & 1 & 4 & 2 & 1 & 1 & 1 & 0 & -1 & 1 & -1 & 1 & 1 & 1 & -1 & 1 & -1 & 1 & $(2,8,1)\bm{2}$ \\
0 & 1 & 3 & 2 & 1 & 1 & 4 & 2 & 1 & 1 & 1 & 0 & -1 & 1 & -1 & 1 & -1 & 1 & -1 & 1 & -1 & 1 & $(2,8,1)\bm{2}$ \\
0 & 1 & 3 & 2 & 1 & 1 & 4 & 2 & 0 & 1 & 2 & 0 & 1 & -1 & -1 & 1 & 1 & -1 & 1 & -1 & -1 & 1 & $(2,8,1)\bm{2}$ \\
\hline
\end{tabularx}
\end{table}

\begin{table}[H]
\caption{$(T_1,T_2)$-charges and coupling coefficients in type $(N_1,N_2,k)=(2,9,\frac{1}{2})$ giving favorable structures.
The last column shows the favorable structures given in each pattern of $(T_1,T_2)$-charges and coupling coefficients.}
\label{tab:T-charges (2,9,1/2)}
\centering
\footnotesize
\begin{tabularx}{180mm}{CCCCCCCCCCCCCCCCCCCCCC|c} \hline
$p_{Q11}$ & $p_{Q21}$ & $p_{Q12}$ & $p_{Q22}$ & $p_{u11}$ & $p_{u21}$ & $p_{u12}$ & $p_{u22}$ & $p_{d11}$ & $p_{d21}$ & $p_{d12}$ & $p_{d22}$ & $a_{22}$ & $a_{23}$ & $a_{32}$ & $a_{33}$ & $b_{21}$ & $b_{22}$ & $b_{23}$ & $b_{31}$ & $b_{32}$ & $b_{33}$ & Label \\ \hline
0 & 1 & 3 & 2 & 0 & 1 & 4 & 2 & 0 & 1 & 1 & 0 & -1 & -1 & -1 & 1 & 1 & 1 & -1 & 1 & 1 & -1 & $(2,9,\frac{1}{2})\bm{1}$ \\
0 & 1 & 3 & 2 & 0 & 1 & 4 & 2 & 0 & 1 & 1 & 0 & -1 & -1 & -1 & 1 & -1 & 1 & -1 & 1 & 1 & -1 & $(2,9,\frac{1}{2})\bm{1}$ \\
0 & 1 & 3 & 2 & 0 & 1 & 4 & 2 & 0 & 1 & 1 & 0 & -1 & -1 & -1 & 1 & 1 & 1 & -1 & -1 & 1 & -1 & $(2,9,\frac{1}{2})\bm{2}$ \\
0 & 1 & 3 & 2 & 0 & 1 & 4 & 2 & 1 & 1 & 1 & 0 & -1 & -1 & -1 & 1 & -1 & 1 & -1 & 1 & 1 & -1 & $(2,9,\frac{1}{2})\bm{3}$ \\
0 & 1 & 3 & 2 & 0 & 1 & 4 & 2 & 1 & 1 & 1 & 0 & -1 & -1 & -1 & 1 & 1 & 1 & -1 & -1 & 1 & -1 & $(2,9,\frac{1}{2})\bm{4}$ \\
0 & 1 & 3 & 2 & 0 & 1 & 4 & 2 & 1 & 1 & 1 & 0 & -1 & -1 & -1 & 1 & -1 & 1 & -1 & -1 & 1 & -1 & $(2,9,\frac{1}{2})\bm{4}$ \\
0 & 1 & 3 & 2 & 0 & 1 & 4 & 2 & 0 & 1 & 1 & 0 & 1 & 1 & 1 & -1 & -1 & -1 & 1 & 1 & -1 & 1 & $(2,9,\frac{1}{2})\bm{5}$ \\
\hline
\end{tabularx}
\end{table}
\begin{table}[H]
\centering
\footnotesize
\begin{tabularx}{180mm}{CCCCCCCCCCCCCCCCCCCCCC|c} \hline
$p_{Q11}$ & $p_{Q21}$ & $p_{Q12}$ & $p_{Q22}$ & $p_{u11}$ & $p_{u21}$ & $p_{u12}$ & $p_{u22}$ & $p_{d11}$ & $p_{d21}$ & $p_{d12}$ & $p_{d22}$ & $a_{22}$ & $a_{23}$ & $a_{32}$ & $a_{33}$ & $b_{21}$ & $b_{22}$ & $b_{23}$ & $b_{31}$ & $b_{32}$ & $b_{33}$ & Label \\ \hline
0 & 1 & 3 & 2 & 0 & 1 & 5 & 2 & 1 & 1 & 1 & 0 & -1 & 1 & -1 & -1 & 1 & 1 & -1 & -1 & 1 & -1 & $(2,9,\frac{1}{2})\bm{15}$ \\
0 & 1 & 3 & 2 & 0 & 1 & 5 & 2 & 1 & 1 & 1 & 0 & 1 & -1 & -1 & -1 & -1 & -1 & 1 & -1 & 1 & -1 & $(2,9,\frac{1}{2})\bm{15}$ \\
0 & 1 & 3 & 2 & 0 & 1 & 5 & 2 & 1 & 1 & 1 & 0 & -1 & -1 & -1 & -1 & -1 & -1 & 1 & -1 & 1 & -1 & $(2,9,\frac{1}{2})\bm{15}$ \\
0 & 1 & 3 & 2 & 1 & 0 & 4 & 1 & 0 & 1 & 1 & 0 & -1 & 1 & -1 & -1 & 1 & 1 & -1 & 1 & 1 & -1 & $(2,9,\frac{1}{2})\bm{16}$ \\
0 & 1 & 3 & 2 & 1 & 0 & 4 & 1 & 0 & 1 & 1 & 0 & 1 & -1 & -1 & -1 & -1 & -1 & 1 & 1 & 1 & -1 & $(2,9,\frac{1}{2})\bm{16}$ \\
0 & 1 & 3 & 2 & 1 & 0 & 4 & 1 & 0 & 1 & 1 & 0 & -1 & 1 & -1 & -1 & 1 & 1 & -1 & -1 & 1 & -1 & $(2,9,\frac{1}{2})\bm{17}$ \\
0 & 1 & 3 & 2 & 1 & 0 & 4 & 1 & 0 & 1 & 1 & 0 & 1 & -1 & -1 & -1 & -1 & -1 & 1 & -1 & 1 & -1 & $(2,9,\frac{1}{2})\bm{17}$ \\
\hline
\end{tabularx}
\end{table}
\begin{table}[H]
\centering
\footnotesize
\begin{tabularx}{180mm}{CCCCCCCCCCCCCCCCCCCCCC|c} \hline
$p_{Q11}$ & $p_{Q21}$ & $p_{Q12}$ & $p_{Q22}$ & $p_{u11}$ & $p_{u21}$ & $p_{u12}$ & $p_{u22}$ & $p_{d11}$ & $p_{d21}$ & $p_{d12}$ & $p_{d22}$ & $a_{22}$ & $a_{23}$ & $a_{32}$ & $a_{33}$ & $b_{21}$ & $b_{22}$ & $b_{23}$ & $b_{31}$ & $b_{32}$ & $b_{33}$ & Label \\ \hline
0 & 1 & 4 & 2 & 0 & 0 & 4 & 2 & 0 & 1 & 1 & 0 & -1 & -1 & -1 & 1 & 1 & 1 & -1 & 1 & 1 & -1 & $(2,9,\frac{1}{2})\bm{26}$ \\
0 & 1 & 4 & 2 & 0 & 0 & 4 & 2 & 0 & 1 & 1 & 0 & -1 & -1 & 1 & -1 & 1 & 1 & -1 & -1 & -1 & 1 & $(2,9,\frac{1}{2})\bm{26}$ \\
1 & 1 & 2 & 3 & 0 & 0 & 5 & 1 & 0 & 1 & 2 & 0 & 1 & -1 & -1 & 1 & 1 & -1 & 1 & 1 & -1 & -1 & $(2,9,\frac{1}{2})\bm{27}$ \\
1 & 1 & 2 & 3 & 0 & 0 & 5 & 1 & 0 & 1 & 2 & 0 & -1 & 1 & 1 & -1 & -1 & 1 & -1 & -1 & 1 & 1 & $(2,9,\frac{1}{2})\bm{27}$ \\
1 & 1 & 2 & 3 & 0 & 0 & 5 & 1 & 0 & 1 & 2 & 0 & 1 & -1 & -1 & 1 & -1 & -1 & 1 & 1 & -1 & -1 & $(2,9,\frac{1}{2})\bm{28}$ \\
1 & 1 & 2 & 3 & 0 & 0 & 5 & 1 & 0 & 1 & 2 & 0 & -1 & 1 & 1 & -1 & 1 & 1 & -1 & -1 & 1 & 1 & $(2,9,\frac{1}{2})\bm{28}$ \\
1 & 1 & 2 & 3 & 0 & 0 & 5 & 1 & 0 & 1 & 2 & 0 & 1 & -1 & -1 & 1 & 1 & 1 & -1 & 1 & -1 & -1 & $(2,9,\frac{1}{2})\bm{29}$ \\
\hline
\end{tabularx}
\end{table}
\begin{table}[H]
\centering
\footnotesize
\begin{tabularx}{180mm}{CCCCCCCCCCCCCCCCCCCCCC|c} \hline
$p_{Q11}$ & $p_{Q21}$ & $p_{Q12}$ & $p_{Q22}$ & $p_{u11}$ & $p_{u21}$ & $p_{u12}$ & $p_{u22}$ & $p_{d11}$ & $p_{d21}$ & $p_{d12}$ & $p_{d22}$ & $a_{22}$ & $a_{23}$ & $a_{32}$ & $a_{33}$ & $b_{21}$ & $b_{22}$ & $b_{23}$ & $b_{31}$ & $b_{32}$ & $b_{33}$ & Label \\ \hline
1 & 1 & 3 & 2 & 0 & 0 & 4 & 2 & 0 & 1 & 1 & 0 & -1 & 1 & -1 & -1 & 1 & 1 & -1 & -1 & 1 & -1 & $(2,9,\frac{1}{2})\bm{42}$ \\
1 & 1 & 3 & 2 & 0 & 0 & 4 & 2 & 0 & 1 & 1 & 0 & 1 & -1 & -1 & -1 & -1 & -1 & 1 & 1 & 1 & -1 & $(2,9,\frac{1}{2})\bm{42}$ \\
1 & 1 & 3 & 2 & 0 & 0 & 4 & 2 & 0 & 1 & 1 & 0 & 1 & -1 & -1 & -1 & -1 & -1 & 1 & -1 & 1 & -1 & $(2,9,\frac{1}{2})\bm{42}$ \\
1 & 1 & 3 & 2 & 0 & 0 & 4 & 2 & 1 & 1 & 2 & 0 & -1 & 1 & -1 & -1 & 1 & -1 & 1 & 1 & -1 & 1 & $(2,9,\frac{1}{2})\bm{43}$ \\
1 & 1 & 3 & 2 & 0 & 0 & 4 & 2 & 1 & 1 & 2 & 0 & -1 & 1 & -1 & -1 & 1 & -1 & 1 & -1 & -1 & 1 & $(2,9,\frac{1}{2})\bm{43}$ \\
1 & 1 & 3 & 2 & 0 & 0 & 4 & 2 & 1 & 1 & 2 & 0 & 1 & -1 & -1 & -1 & -1 & 1 & -1 & 1 & -1 & 1 & $(2,9,\frac{1}{2})\bm{43}$ \\
1 & 1 & 3 & 2 & 0 & 0 & 4 & 2 & 1 & 1 & 2 & 0 & 1 & -1 & -1 & -1 & -1 & 1 & -1 & -1 & -1 & 1 & $(2,9,\frac{1}{2})\bm{43}$ \\
\hline
\end{tabularx}
\end{table}
\begin{table}[H]
\centering
\footnotesize
\begin{tabularx}{180mm}{CCCCCCCCCCCCCCCCCCCCCC|c} \hline
$p_{Q11}$ & $p_{Q21}$ & $p_{Q12}$ & $p_{Q22}$ & $p_{u11}$ & $p_{u21}$ & $p_{u12}$ & $p_{u22}$ & $p_{d11}$ & $p_{d21}$ & $p_{d12}$ & $p_{d22}$ & $a_{22}$ & $a_{23}$ & $a_{32}$ & $a_{33}$ & $b_{21}$ & $b_{22}$ & $b_{23}$ & $b_{31}$ & $b_{32}$ & $b_{33}$ & Label \\ \hline
1 & 1 & 3 & 2 & 1 & 0 & 4 & 1 & 1 & 1 & 1 & 0 & -1 & 1 & 1 & 1 & 1 & -1 & 1 & -1 & 1 & -1 & $(2,9,\frac{1}{2})\bm{53}$ \\
1 & 1 & 3 & 2 & 1 & 0 & 4 & 1 & 1 & 1 & 1 & 0 & 1 & -1 & 1 & 1 & -1 & 1 & -1 & -1 & 1 & -1 & $(2,9,\frac{1}{2})\bm{53}$ \\
1 & 1 & 3 & 2 & 1 & 0 & 4 & 1 & 1 & 1 & 1 & 0 & -1 & 1 & -1 & -1 & 1 & -1 & 1 & 1 & -1 & 1 & $(2,9,\frac{1}{2})\bm{53}$ \\
1 & 1 & 3 & 2 & 1 & 0 & 4 & 1 & 1 & 1 & 1 & 0 & 1 & -1 & -1 & -1 & -1 & 1 & -1 & 1 & -1 & 1 & $(2,9,\frac{1}{2})\bm{53}$ \\
1 & 1 & 3 & 2 & 1 & 0 & 4 & 1 & 1 & 1 & 1 & 0 & -1 & 1 & 1 & 1 & -1 & -1 & 1 & -1 & 1 & -1 & $(2,9,\frac{1}{2})\bm{54}$ \\
1 & 1 & 3 & 2 & 1 & 0 & 4 & 1 & 1 & 1 & 1 & 0 & 1 & -1 & 1 & 1 & 1 & 1 & -1 & -1 & 1 & -1 & $(2,9,\frac{1}{2})\bm{54}$ \\
1 & 1 & 3 & 2 & 1 & 0 & 4 & 1 & 1 & 1 & 1 & 0 & -1 & 1 & -1 & -1 & -1 & -1 & 1 & 1 & -1 & 1 & $(2,9,\frac{1}{2})\bm{54}$ \\
\hline
\end{tabularx}
\end{table}
\begin{table}[H]
\centering
\footnotesize
\begin{tabularx}{180mm}{CCCCCCCCCCCCCCCCCCCCCC|c} \hline
$p_{Q11}$ & $p_{Q21}$ & $p_{Q12}$ & $p_{Q22}$ & $p_{u11}$ & $p_{u21}$ & $p_{u12}$ & $p_{u22}$ & $p_{d11}$ & $p_{d21}$ & $p_{d12}$ & $p_{d22}$ & $a_{22}$ & $a_{23}$ & $a_{32}$ & $a_{33}$ & $b_{21}$ & $b_{22}$ & $b_{23}$ & $b_{31}$ & $b_{32}$ & $b_{33}$ & Label \\ \hline
1 & 1 & 3 & 2 & 1 & 0 & 5 & 2 & 1 & 1 & 2 & 0 & 1 & -1 & -1 & -1 & -1 & 1 & 1 & -1 & -1 & -1 & $(2,9,\frac{1}{2})\bm{57}$ \\
1 & 1 & 3 & 2 & 1 & 0 & 5 & 2 & 1 & 1 & 2 & 0 & -1 & 1 & 1 & 1 & 1 & 1 & -1 & 1 & -1 & 1 & $(2,9,\frac{1}{2})\bm{58}$ \\
1 & 1 & 3 & 2 & 1 & 0 & 5 & 2 & 1 & 1 & 2 & 0 & 1 & -1 & 1 & 1 & -1 & -1 & 1 & 1 & -1 & 1 & $(2,9,\frac{1}{2})\bm{58}$ \\
1 & 1 & 3 & 2 & 1 & 0 & 5 & 2 & 1 & 1 & 2 & 0 & -1 & 1 & -1 & -1 & 1 & 1 & -1 & -1 & 1 & -1 & $(2,9,\frac{1}{2})\bm{58}$ \\
1 & 1 & 3 & 2 & 1 & 0 & 5 & 2 & 1 & 1 & 2 & 0 & 1 & -1 & -1 & -1 & -1 & -1 & 1 & -1 & 1 & -1 & $(2,9,\frac{1}{2})\bm{58}$ \\
1 & 1 & 3 & 2 & 1 & 0 & 5 & 2 & 1 & 1 & 2 & 0 & -1 & 1 & 1 & 1 & 1 & -1 & 1 & 1 & 1 & -1 & $(2,9,\frac{1}{2})\bm{59}$ \\
1 & 1 & 3 & 2 & 1 & 0 & 5 & 2 & 1 & 1 & 2 & 0 & -1 & 1 & 1 & 1 & 1 & -1 & 1 & -1 & 1 & -1 & $(2,9,\frac{1}{2})\bm{59}$ \\
\hline
\end{tabularx}
\end{table}

\begin{table}[H]
\caption{$(T_1,T_2)$-charges and coupling coefficients in type $(N_1,N_2,k)=(2,9,1)$ giving favorable structures.
The last column shows the favorable structures given in each pattern of $(T_1,T_2)$-charges and coupling coefficients.}
\label{tab:T-charges (2,9,1)}
\centering
\footnotesize
\begin{tabularx}{180mm}{CCCCCCCCCCCCCCCCCCCCCC|c} \hline
$p_{Q11}$ & $p_{Q21}$ & $p_{Q12}$ & $p_{Q22}$ & $p_{u11}$ & $p_{u21}$ & $p_{u12}$ & $p_{u22}$ & $p_{d11}$ & $p_{d21}$ & $p_{d12}$ & $p_{d22}$ & $a_{22}$ & $a_{23}$ & $a_{32}$ & $a_{33}$ & $b_{21}$ & $b_{22}$ & $b_{23}$ & $b_{31}$ & $b_{32}$ & $b_{33}$ & Label \\ \hline
0 & 1 & 3 & 2 & 0 & 1 & 5 & 2 & 0 & 1 & 1 & 0 & -1 & 1 & 1 & -1 & 1 & 1 & -1 & -1 & 1 & -1 & $(2,9,1)\bm{1}$ \\
0 & 1 & 3 & 2 & 0 & 1 & 5 & 2 & 0 & 1 & 1 & 0 & 1 & -1 & 1 & -1 & -1 & -1 & 1 & -1 & 1 & -1 & $(2,9,1)\bm{1}$ \\
0 & 1 & 3 & 2 & 1 & 1 & 4 & 2 & 0 & 1 & 1 & 0 & -1 & 1 & -1 & 1 & 1 & 1 & -1 & 1 & -1 & 1 & $(2,9,1)\bm{1}$ \\
0 & 1 & 3 & 2 & 1 & 1 & 4 & 2 & 0 & 1 & 1 & 0 & 1 & -1 & -1 & 1 & -1 & -1 & 1 & 1 & -1 & 1 & $(2,9,1)\bm{1}$ \\
0 & 1 & 3 & 2 & 0 & 1 & 5 & 2 & 0 & 1 & 2 & 0 & -1 & 1 & 1 & -1 & 1 & 1 & -1 & 1 & 1 & -1 & $(2,9,1)\bm{2}$ \\
0 & 1 & 3 & 2 & 0 & 1 & 5 & 2 & 0 & 1 & 2 & 0 & -1 & 1 & 1 & -1 & -1 & 1 & -1 & 1 & 1 & -1 & $(2,9,1)\bm{2}$ \\
0 & 1 & 3 & 2 & 0 & 1 & 5 & 2 & 1 & 1 & 1 & 0 & -1 & 1 & 1 & -1 & 1 & 1 & -1 & -1 & 1 & -1 & $(2,9,1)\bm{2}$ \\
\hline
\end{tabularx}
\end{table}
\begin{table}[H]
\centering
\footnotesize
\begin{tabularx}{180mm}{CCCCCCCCCCCCCCCCCCCCCC|c} \hline
$p_{Q11}$ & $p_{Q21}$ & $p_{Q12}$ & $p_{Q22}$ & $p_{u11}$ & $p_{u21}$ & $p_{u12}$ & $p_{u22}$ & $p_{d11}$ & $p_{d21}$ & $p_{d12}$ & $p_{d22}$ & $a_{22}$ & $a_{23}$ & $a_{32}$ & $a_{33}$ & $b_{21}$ & $b_{22}$ & $b_{23}$ & $b_{31}$ & $b_{32}$ & $b_{33}$ & Label \\ \hline
1 & 1 & 3 & 2 & 1 & 0 & 5 & 1 & 0 & 1 & 1 & 0 & 1 & -1 & 1 & 1 & -1 & -1 & 1 & -1 & -1 & 1 & $(2,9,1)\bm{13}$ \\
1 & 1 & 3 & 2 & 1 & 0 & 5 & 1 & 0 & 1 & 1 & 0 & -1 & 1 & -1 & -1 & 1 & 1 & -1 & 1 & 1 & -1 & $(2,9,1)\bm{13}$ \\
1 & 1 & 3 & 2 & 1 & 0 & 5 & 1 & 0 & 1 & 1 & 0 & 1 & -1 & -1 & -1 & -1 & -1 & 1 & 1 & 1 & -1 & $(2,9,1)\bm{13}$ \\
1 & 1 & 3 & 2 & 1 & 0 & 5 & 1 & 0 & 1 & 1 & 0 & -1 & 1 & 1 & 1 & -1 & 1 & -1 & -1 & -1 & 1 & $(2,9,1)\bm{14}$ \\
1 & 1 & 3 & 2 & 1 & 0 & 5 & 1 & 0 & 1 & 1 & 0 & 1 & -1 & 1 & 1 & 1 & -1 & 1 & -1 & -1 & 1 & $(2,9,1)\bm{14}$ \\
1 & 1 & 3 & 2 & 1 & 0 & 5 & 1 & 0 & 1 & 1 & 0 & -1 & 1 & -1 & -1 & -1 & 1 & -1 & 1 & 1 & -1 & $(2,9,1)\bm{14}$ \\
1 & 1 & 3 & 2 & 1 & 0 & 5 & 1 & 0 & 1 & 1 & 0 & 1 & -1 & -1 & -1 & 1 & -1 & 1 & 1 & 1 & -1 & $(2,9,1)\bm{14}$ \\
\hline
\end{tabularx}
\end{table}

\begin{table}[H]
\caption{$(T_1,T_2)$-charges and coupling coefficients in type $(N_1,N_2,k)=(3,6,\frac{1}{2})$ giving favorable structures.
The last column shows the favorable structures given in each pattern of $(T_1,T_2)$-charges and coupling coefficients.}
\label{tab:T-charges (3,6,1/2)}
\centering
\footnotesize
\begin{tabularx}{180mm}{CCCCCCCCCCCCCCCCCCCCCC|c} \hline
$p_{Q11}$ & $p_{Q21}$ & $p_{Q12}$ & $p_{Q22}$ & $p_{u11}$ & $p_{u21}$ & $p_{u12}$ & $p_{u22}$ & $p_{d11}$ & $p_{d21}$ & $p_{d12}$ & $p_{d22}$ & $a_{22}$ & $a_{23}$ & $a_{32}$ & $a_{33}$ & $b_{21}$ & $b_{22}$ & $b_{23}$ & $b_{31}$ & $b_{32}$ & $b_{33}$ & Label \\ \hline
1 & 2 & 2 & 1 & 1 & 1 & 3 & 2 & 0 & 2 & 1 & 0 & 1 & 1 & -1 & 1 & 1 & 1 & -1 & 1 & 1 & 1 & $(3,6,\frac{1}{2})\bm{1}$ \\
1 & 2 & 2 & 1 & 1 & 1 & 3 & 2 & 0 & 2 & 1 & 0 & 1 & 1 & -1 & 1 & 1 & 1 & -1 & -1 & 1 & 1 & $(3,6,\frac{1}{2})\bm{1}$ \\
1 & 2 & 2 & 1 & 1 & 1 & 3 & 2 & 0 & 2 & 1 & 0 & -1 & 1 & -1 & 1 & 1 & -1 & -1 & 1 & -1 & 1 & $(3,6,\frac{1}{2})\bm{1}$ \\
1 & 2 & 2 & 1 & 1 & 1 & 3 & 2 & 0 & 2 & 1 & 0 & -1 & 1 & -1 & 1 & 1 & -1 & -1 & -1 & -1 & 1 & $(3,6,\frac{1}{2})\bm{1}$ \\
1 & 2 & 2 & 1 & 1 & 1 & 3 & 2 & 0 & 2 & 1 & 0 & 1 & -1 & -1 & 1 & -1 & 1 & 1 & 1 & -1 & 1 & $(3,6,\frac{1}{2})\bm{1}$ \\
1 & 2 & 2 & 1 & 1 & 1 & 3 & 2 & 0 & 2 & 1 & 0 & 1 & -1 & -1 & 1 & -1 & 1 & 1 & -1 & -1 & 1 & $(3,6,\frac{1}{2})\bm{1}$ \\
1 & 2 & 2 & 1 & 1 & 1 & 3 & 2 & 0 & 2 & 1 & 0 & -1 & -1 & -1 & 1 & -1 & -1 & 1 & 1 & 1 & 1 & $(3,6,\frac{1}{2})\bm{1}$ \\
\hline
\end{tabularx}
\end{table}

\begin{table}[H]
\caption{$(T_1,T_2)$-charges and coupling coefficients in type $(N_1,N_2,k)=(3,6,2)$ giving favorable structures.
The last column shows the favorable structures given in each pattern of $(T_1,T_2)$-charges and coupling coefficients.}
\label{tab:T-charges (3,6,2)}
\centering
\footnotesize
\begin{tabularx}{180mm}{CCCCCCCCCCCCCCCCCCCCCC|c} \hline
$p_{Q11}$ & $p_{Q21}$ & $p_{Q12}$ & $p_{Q22}$ & $p_{u11}$ & $p_{u21}$ & $p_{u12}$ & $p_{u22}$ & $p_{d11}$ & $p_{d21}$ & $p_{d12}$ & $p_{d22}$ & $a_{22}$ & $a_{23}$ & $a_{32}$ & $a_{33}$ & $b_{21}$ & $b_{22}$ & $b_{23}$ & $b_{31}$ & $b_{32}$ & $b_{33}$ & Label \\ \hline
0 & 1 & 3 & 0 & 1 & 2 & 2 & 4 & 0 & 0 & 1 & 0 & -1 & -1 & 1 & 1 & 1 & -1 & 1 & 1 & -1 & -1 & $(3,6,2)\bm{1}$ \\
0 & 1 & 3 & 0 & 1 & 2 & 2 & 4 & 0 & 0 & 1 & 0 & -1 & -1 & 1 & 1 & 1 & 1 & -1 & 1 & -1 & -1 & $(3,6,2)\bm{1}$ \\
0 & 1 & 3 & 0 & 1 & 2 & 2 & 4 & 0 & 0 & 1 & 0 & -1 & -1 & -1 & 1 & 1 & -1 & 1 & 1 & -1 & -1 & $(3,6,2)\bm{1}$ \\
0 & 1 & 3 & 0 & 1 & 2 & 2 & 4 & 0 & 0 & 1 & 0 & -1 & -1 & -1 & 1 & 1 & 1 & -1 & 1 & -1 & -1 & $(3,6,2)\bm{1}$ \\
0 & 1 & 3 & 0 & 1 & 2 & 2 & 4 & 0 & 0 & 1 & 0 & -1 & -1 & 1 & -1 & 1 & -1 & 1 & -1 & 1 & 1 & $(3,6,2)\bm{1}$ \\
0 & 1 & 3 & 0 & 1 & 2 & 2 & 4 & 0 & 0 & 1 & 0 & -1 & -1 & 1 & -1 & 1 & 1 & -1 & -1 & 1 & 1 & $(3,6,2)\bm{1}$ \\
0 & 1 & 3 & 0 & 1 & 2 & 2 & 4 & 0 & 0 & 1 & 0 & -1 & -1 & -1 & -1 & 1 & -1 & 1 & -1 & 1 & 1 & $(3,6,2)\bm{1}$ \\
\hline
\end{tabularx}
\end{table}

\clearpage

\end{document}